\documentclass[useAMS]{mn2e}
\usepackage{amsmath}
\usepackage{times}
\usepackage{graphicx}

\newcommand{\Teff}{\ensuremath{T_{\rm eff}}}                            % Effective temperature symbol
\newcommand{\logg}{\ensuremath{\log g}}                                 % log(g) symbol
\newcommand{\Msun}{\ensuremath{\,{\rm M}_\odot}}                        % Solar mass symbol
\newcommand{\Rsun}{\ensuremath{\,{\rm R}_\odot}}                        % Solar radius symbol
\newcommand{\ion}[2]{{#1}\,{\sc {\small{#2}}}}                          % This creates ion symbols using small caps
\newcommand{\kms}{\,km\,s$^{-1}$}                                       % km/s symbol
\newcommand{\spd}{{\sc spd}}

\title[]{Spectroscopically resolving the Algol triple system}

\author[V.\ Kolbas et al.]
       {V.~Kolbas$^1$, K.~Pavlovski$^1$, J.~Southworth$^2$, C.-U.~Lee$^3$, D.-J. Lee$^3$,
        J.~W.~Lee$^3$,
        \newauthor
        S.-L.~Kim$^3$, H.-I.~Kim$^3$, B.~Smalley$^2$ and A.~Tkachenko$^4$ \\
        $^1$\,Department of Physics, University of Zagreb, Bijeni\v{c}ka cesta 32,
        10000 Zagreb, Croatia \\
        $^2$\,Astrophysics Group, Keele University, Staffordshire, ST5 5BG, UK \\
        $^3$\,Korea Astronomy and Space Science Institute, 776 Daedukdae-ro, Yuseong-gu,
             Daejeon 305-348, Republic of Korea \\  %%%  South Korea \\
        $^4$\,Instituut voor Sterrenkunde, KU Leuven, Celestijnenlaan 200D, B-3001 Leuven,
        Belgium \\
}

\date{}

\pagerange{\pageref{firstpage}--\pageref{lastpage}} \pubyear{2014}

\begin{document} \maketitle \label{firstpage} %%%%%%%%%%%%%%%%%%%%%%%%%%%%%%%%%%%%%%%%%%%%%%%%%%%%%%%%%%%%%%%%%%%%%%%%%%%%%%%%%%%%%%%%%%%%%%%%%%%%%%%
%%%%%%%%%%%%%%%%%%%%%%%%%%%%%%%%%%%%%%%%%%%%%%%%%%%%%%%%%%%%%%%%%%%%%%%%%%%%%%%%%%%%%%%%%%%%%%%%%%%%%%%%%%%%%%%%%%%%%%%%%%%%%%%%%%%%%%%%%%%%%%%%%%%%%

\begin{abstract}
Algol ($\beta$\,Persei) is the prototypical semi-detached eclipsing binary and a hierarchical 
triple system. From 2006 to 2010 we obtained 121 high-resolution and high-S/N \'{e}chelle 
spectra of this object. Spectral disentangling yields the individual spectra of all three 
stars, and greatly improved elements both the inner and outer orbits. We find masses of 
$M_{\rm A} = 3.39\pm0.06$\Msun, $M_{\rm B} = 0.770\pm0.009$\Msun\ and $M_{\rm C} = 
1.58\pm0.09$\Msun. The disentangled spectra also give the light ratios between the 
components in the $B$ and $V$ bands. Atmospheric parameters for the three stars are 
determined, including detailed elemental abundances for Algol A and Algol C. We find 
the following effective temperatures: $T_{\rm A} = 12\,550\pm120$\,K, $T_{\rm B} = 
4900\pm300$\,K and $T_{\rm C} = 7550\pm250$\,K. The projected rotational velocities 
are $v_{\rm A} \sin i_{\rm A} = 50.8\pm0.8$\,\kms, $v_{\rm B} \sin i_{\rm B} = 62\pm2$\kms\ 
and $v_{\rm C} \sin i_{\rm C} = 12.4\pm0.6$\kms. This is the first measurement of the 
rotational velocity for Algol B, and confirms that it is synchronous with the orbital motion. 
The abundance patterns of components A and C are identical to within the measurement errors, 
and are basically solar. They can be summarised as mean metal abundances: [M/H]$_{\rm A} = 
-0.03\pm0.08$ and [M/H]$_{\rm C} = 0.04\pm0.09$. A carbon deficiency is confirmed for Algol A, 
with tentative indications for a slight overabundance of nitrogen. The ratio of their 
abundances is (C/N)$_{\rm A} = 2.0\pm0.4$, half of the solar value of (C/N)$_{\odot} = 
4.0\pm0.7$.  The new results derived in this study, including detailed abundances 
and metallicities, will enable tight constraints on theoretical evolutionary models for 
this complex system.
\end{abstract}


%%%%%%%%%%%%%%%%%%%%%%%%%%%%%%%%%%%%%%%%%%%%%%%%%%%%%%%%%%%%%%%%%%%%%%%%%%%%%%%%%%%%%%%%%%%%%%%%%%%%%%%%%%%%%%%%%%%%%%%%%%%%%%%%%%%%%%%%%%%%%%%%%%%%%

\section{Introduction}                                           \label{sec:intro}

In his opening speach to IAU Colloquium 107 (`Algols'), Batten (1989) was tempted to define 
an Algol system simply as ``a binary in which the less massive component fills its Roche lobe  
and the other, which does not, is not degenerate''. The resulting evolutionary paradox can be
 solved by postulating an episode of mass transfer from what was initially the more massive 
component to its then less massive, and consequently less evolved, companion (Crawford 1955). 
This first, rapid, phase of mass exchange between the components eventually leads to an 
Algol-type system. Numerical calculations of the evolution of stars in binary systems have 
proved this hypothesis to be very plausible (cf.\ Paczy\'{n}ski 1971).

Direct comparison of the observed properties of Algol systems to evolutionary model calculations 
are sparse, because only a few Algols have reliable measurements of their physical properties 
(Maxted \& Hilditch 1996), and because models are unavailable for the relevant parameters. 
The main observational obstacle in the determination of the accurate 
stellar quantities for Algols arises due to the intrinsic faintness of the 
Roche-lobe filling mass-losing component, the reason they were detected 
only due to technological advances in light detectors (c.f.~Tomkin 1989).
Model calculations for Algol with both conservative and non-conservative binary evolution 
have only been calculated by Sarna (1993).

Changes in the chemical composition of the surface layers of stars which have experienced mass 
transfer are a feature of theoretical predictions. Layers that were once deep inside the star 
become exposed after mass transfer. This makes it possible to observationally probe CNO 
nucleosynthetic processes in stellar cores, and the efficiency of different mixing processes 
in stellar interiors. In that sense CNO abundances can serve as a sensitive probe of stellar 
structure and evolution.

Carbon depletions have been detected in Algol and in Algol-type systems (Parthasarathy, Lambert 
\& Tomkin 1979, 1983; Cugier \& Hardorp 1988; Cugier 1989; Tomkin et al.\ 1993, Ibano\u{g}lu et 
al.\ 2012). However, comparisons to theoretical models were rather scarce since detailed 
evolutionary calculations were performed for only a few Algol-type systems, and were limited 
to tracing the evolution of carbon (De Greve \& Cugier 1989; Sarna 1992, 1993; De Greve 1993). 
On the observational side, Algol systems can be rather difficult. The secondary stars are 
Roche-lobe filling subgiants which are usually intrinsically faint, but still bright enough 
to contaminate the spectra of the primary stars. Many contain third components, as is the 
case for Algol itself, which further complicates spectroscopic analyses. Therefore, we 
initiated a new observational project to obtain high-resolution \'{e}chelle spectroscopy 
of Algol systems to which we could apply the powerful technique of spectral disentangling 
(hereafter \spd) and thus obtain the individual spectra of the component stars. This makes 
detailed spectroscopic analyses much more straightforward, and allows the study of the 
spectral characteristics of all components over wide wavelength ranges. To aid the comparison 
with theoretical evolutionary models of semi-detached binary systems, we have also considered 
other diagnostic species. The primary target is nitrogen abundance, which is very sensitive 
to CNO nucleosynthetic processes. Theoretical models of binary star evolution have also been 
calculated for non-conservative mass transfer. A detailed description of our methodology was 
given in a study of the hot Algol u\,Her (Kolbas et al.\ 2014). %Only a systematic study of 
a large sample of mass transferring binaries of Algol type can reveal all the different 
evolutionary paths which these complex binary systems encounter.

The chemical composition was determined from spectra covering the full optical range for 
only a few Algol-type binaries. Beside u\,Her (Kolbas et al.\ 2014), abundance patterns 
have been derived for RZ\,Cas (Narusawa et al.\ 2006, Tkachenko et al.\ 2009), TW\,Dra 
(Tkachenko et al.\ 2010), TX\,UMa (Glazunova et al.\ 2011), and AS\,Eri (Narusawa 2013). 
All these objects except u\,Her have a pulsating primary component.

 In this work we present new high-quality optical spectroscopy of Algol with two main 
goals Firstly, improved measurement of the masses of all three components through spectral 
disentangling, with a special emphasis in studying the spectrum of a faint Roche-lobe-filling 
component over a broad spectral range. Secondly, spectroscopic determination of the effective 
temperature, abundance patterns and bulk metallicities from disentangled spectra of the 
individual components. Our aim is to assemble useful observational constraints for comparison 
to evolutionary model calculations; in particular, the first measurement of the bulk metallicity 
of the components of Algol.

We first summarise the rich observational history of Algol, and highlight some recent 
observational results. In Sect.~3 we present new \'{e}chelle spectra of Algol. Sect.~4 covers 
the \spd\ of the spectra and the determination of new orbital elements for both inner and outer
 orbits. The analysis of the disentangled spectra for all three components is presented in 
Sect.\ 5, including detailed calculations of abundances for Algol A and metallicity for Algol 
C. The results are discussed in the context of mass transfer between Algol A and B in Sect.~6. 
Finally, Sect.~7 summarises our results and discusses future work.

%%%%%%%%%%%%%%%%%%%%%%%%%%%%%%%%%%%%%%%%%%%%%%%%%%%%%%%%%%%%%%%%%%%%%%%%%%%%%%%%%%%%%%%%%%%%%%%%%%%%%%%%%%%%%%%%%%%%%%%%%%%%%%%%%%%%%%%%%%%%%%%%%%%%%

\section{Algol in a nutshell}

Algol ($\beta$\,Persei, HD\,19356) is a hierarchical triple star system (Frieboes-Conde et al.\ 
1970; S\"{o}derhjelm 1980). It is extremely bright ($V = 2.12\,{\rm mag}$) and its periodic 
variability is the subject of a long observational history. There is evidence that Algol's 
periodicity was recorded by the ancient Egyptians three millennia ago, in the Cairo Calendar 
(Jetsu et al.\ 2013). The first well-documented discovery of the 2.867\,day periodicity, and 
an explanation in terms of stellar eclipses, was given by Goodricke (1783) based on naked-eye 
observations. Thanks to its brightness, Algol also has a long history of observations with many 
techniques and at many wavelengths. An outstanding and exhaustive account of these studies was 
given by Wecht (2006).

The inner pair consists of a late-B-type star in orbit with an early-K-type subgiant which 
fills its Roche lobe. This close system exhibits partial eclipses, and is the prototype of 
the Algol class of eclipsing binaries. A tertiary component pursues a 680\,d orbit around 
the inner pair; its spectral classification has been variously given as late-A, early-F, 
and Am. This description of Algol emerged from decades of primarily spectroscopic and 
photometric studies (cf.\ Friebos-Conde et al.\ 1970; Hill et al.\ 1971; S\"{o}derhjelm 1980; 
Richards et al.\ 1988). The main difficulty of these early studies was the intrinsic faintness 
of the cool subgiant, which is exacerbated by dilution due to the light of component C. 
Algol C is brighter than Algol B, but its contribution has previously been very difficult 
to quantify (Richards et al.\ 1988). A breakthrough came with its eventual spectroscopic 
detection in the infrared (Glusheva \& Esipov 1967), and then radial velocity (RV) 
measurements which led to the determination of the dynamical masses for the components 
(Tomkin \& Lambert 1978).

Other observational techniques have been helpful in constraining the orbital and physical 
characteristics of Algol. Rudy \& Kemp (1978) found phase-locked polarization and 
independently determined the inclination of the eclipsing binary. Subsequently Kemp 
et al.\ (1983) discovered eclipse polarization, or the Chandrasekhar effect, further 
evaluated by Wilson \& Liou (1993). The problem of the determination of the mutual 
orientations of the orbits in the Algol system has remainined open (Kemp et al.\ 1981).

Since the successful detection of Algol C with speckle interferometry (Labeyrie et al.\ 
1974), Algol has often been a target for interferometric measurements, which have recently 
settled the issue of the true orientation of both orbits. Csizmadia et al.\ (2009) spatially 
resolved the inner pair with long baseline interferometry in combination with very long 
baseline interferometry (VLBI) radio observations. They found the inner orbit to be prograde, 
in disagreement with the retrograde movement found by Lestrade et al.\ (1993) from radio 
observations. This long-term controversy was solved by Zavala et al.\ (2010), who 
simultaneously resolved all three stellar components in the optical, achieving the 
then highest precision in angular measurements. The outer orbit was shown to be prograde 
and the inner orbit retrograde. VLBA radio (Peterson et al.\ 2010, 2011) and CHARA $H$-band 
measurements (Baron et al.\ 2012) subsequently confirmed these orientations for both orbits. 
The most recent highlight of the interferometric studies is an unambiguous spatial resolution 
of the three stars in the Algol system with angular resolution $<$0.5\,mas (Baron et al.\ 2012). 
This made possible the determination of the orbital and physical characteristics of the components 
(angular sizes and mass ratios) independently from previous studies. Baron et al.\ (2012) also 
determined the mutual inclination of the orbits to be much closer to perpendicularity than 
previously established. The distortion of the Roche-lobe-filling component B is clearly seen 
in the reconstructed image.

Another highlight of recent work on Algol is the first three-dimensional reconstruction of this 
system (Richards et al.\ 2012). These three-dimensional tomograms revealed previously 
undetected evidence of the mass transfer process, such as loop prominences and coronal mass 
ejections. Early predictions of the superhump phenomenon in Algol, i.e.\ the gas between 
the stars in close pair being threaded with a magnetic field even though the hot mass-gaining 
star is not know to have a magnetic field (Retter et al.\ 2005) have been supported by this 
new technique. Algol B is a late-type magnetically active subgiant found to be a strong radio 
and X-ray source (Wade \& Hjellming 1972; White et al.\ 1986; Stern et al.\ 1992, 1995), and 
dominates in the spectrum of Algol at these wavelengths. In particular, X-ray studies have 
made possible a determination of the chemical abundances in the corona of Algol B (Drake 2003) 
as well both the X-ray-bright star Algol B and the X-ray-faint star Algol A (Yang et al.\ 2011).

\begin{figure}
\includegraphics[width=8.5cm]{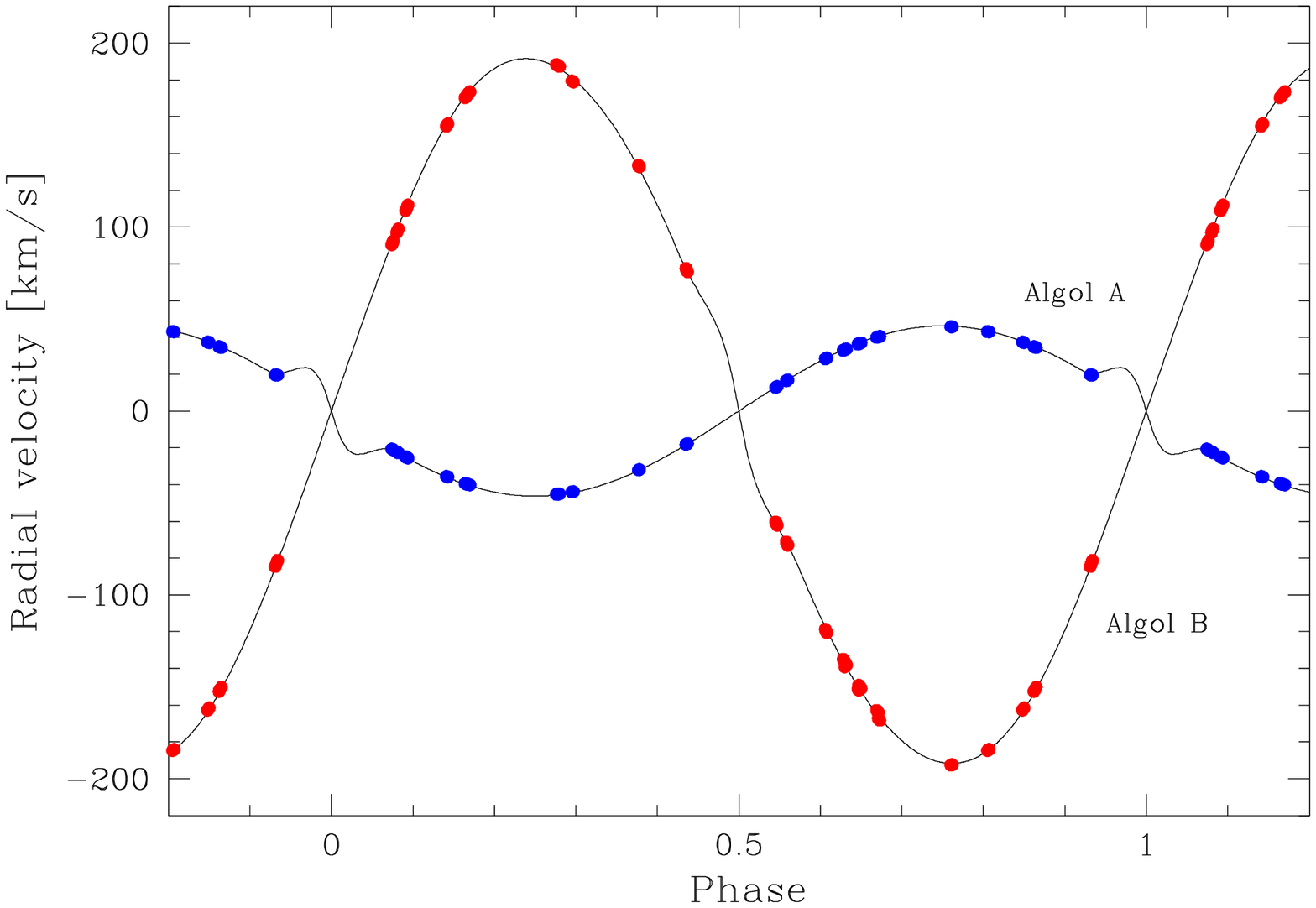}
\includegraphics[width=8.5cm]{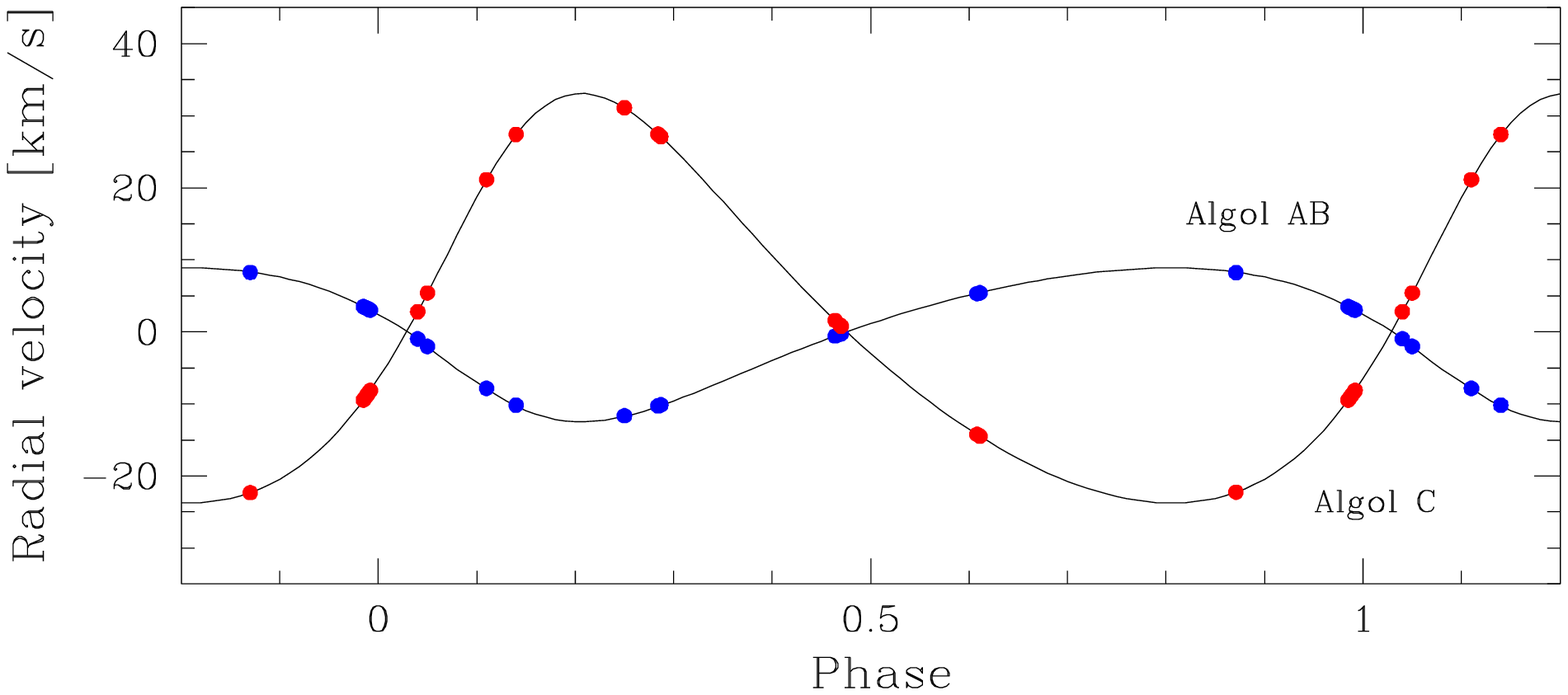}
\caption{ Depiction of the phase distribution of the observed spectra for the
inner (upper panel) and outer (lower panel) systems. The RV curves (solid lines)
were calculated from the orbital elements found via \spd\ (Table~5). Filled circles
represent the RV expected for each observed spectrum, and are not measured RVs.
This is because the \spd\ analysis returns the orbital elements directly, without
passing through the intermediate step of RV determination (cf.\ Simon \& Sturm 1994).
Spectra obtained during eclipse are not shown due to the rotational distortion seen
in the primary eclipse (cf.\ Sect.\ 4). Phases for the inner eclipsing system were
calculated with the ephemeris of Kreiner (2004) with the primary minimum at phase 0.
For the outer system the ephemeris from Baron et al.\ (2012) was used, where
$T_{\rm 0}$ is the time of periastron passage.}
\end{figure}

%%%%%%%%%%%%%%%%%%%%%%%%%%%%%%%%%%%%%%%%%%%%%%%%%%%%%%%%%%%%%%%%%%%%%%%%%%%%%%%%%%%%%%%%%%%%%%%%%%%%%%%%%%%%%%%%%%%%%%%%%%%%%%%%%%%%%%%%%%%%%%%%%%%%%

\section{High-resolution spectroscopy}               \label{sec:obs}

Our observational programme of high-resolution and high signal-to-noise (S/N) spectroscopy 
of Algol was initiated with two observing runs in 2006 and 2007 at the 2.5-m Nordic Optical 
Telescope (NOT) at La Palma, Spain. We obtained 85 spectra of Algol using the Fibre-fed 
Echelle Spectrograph (FIES; Telting et al.\ 2014). FIES is housed in a dedicated climate-controlled
 building and has a high thermal and mechanical stability. The wavelength scale was established 
from thorium-argon exposures taken regularly throughout the observing nights. We used fibre 4 
in bundle B, giving complete spectral coverage in the interval 3640--7360\,\AA\ at a reciprocal 
dispersion ranging from 0.023\,\AA\,px$^{-1}$ in the blue to 0.045\,\AA\,px$^{-1}$ in the red. 
The resolution of the instrument is roughly 3.5\,px, giving a resolving power of 48\,000. 
An exposure time of 15\,s was used for all spectra, resulting in continuum S/N ratios in 
the region of 200--500 in the $B$ and $V$ bands.

In order to cover the long-period outer orbit of Algol, spectroscopic observations were 
taken from 2009 to 2010 at the Bohyunsan Optical Astronomy Observatory (BOAO), South Korea. 
A set of 36 spectra were secured with Bohyunsan Optical \'Echelle Spectrograph (BOES) 
mounted on the 1.8-m telescope (Kim et al.\ 2007). BOES has multiple spectral resolving 
powers up to 80\,000, and covers thee wavelength range 3600 to 10\,200\,\AA. A thorium-argon 
lamp was used for wavelength calibration and S/N ratios of 300--550 were achieved.

The spectra were bias-subtracted, flat-fielded and extracted with the 
{\sc iraf}\footnote{{\sc iraf} is distributed by the National Optical Astronomy Observatory, 
which are operated by the Association of the Universities for Research in Astronomy, Inc., 
under cooperative agreement with the NSF.} \'echelle package routines. Normalisation and 
merging of the orders was performed with great care, using custom programs, to ensure that 
these steps did not cause any systematic errors in the resulting spectra (see Appendix).

\begin{figure*}
\includegraphics[width=18cm]{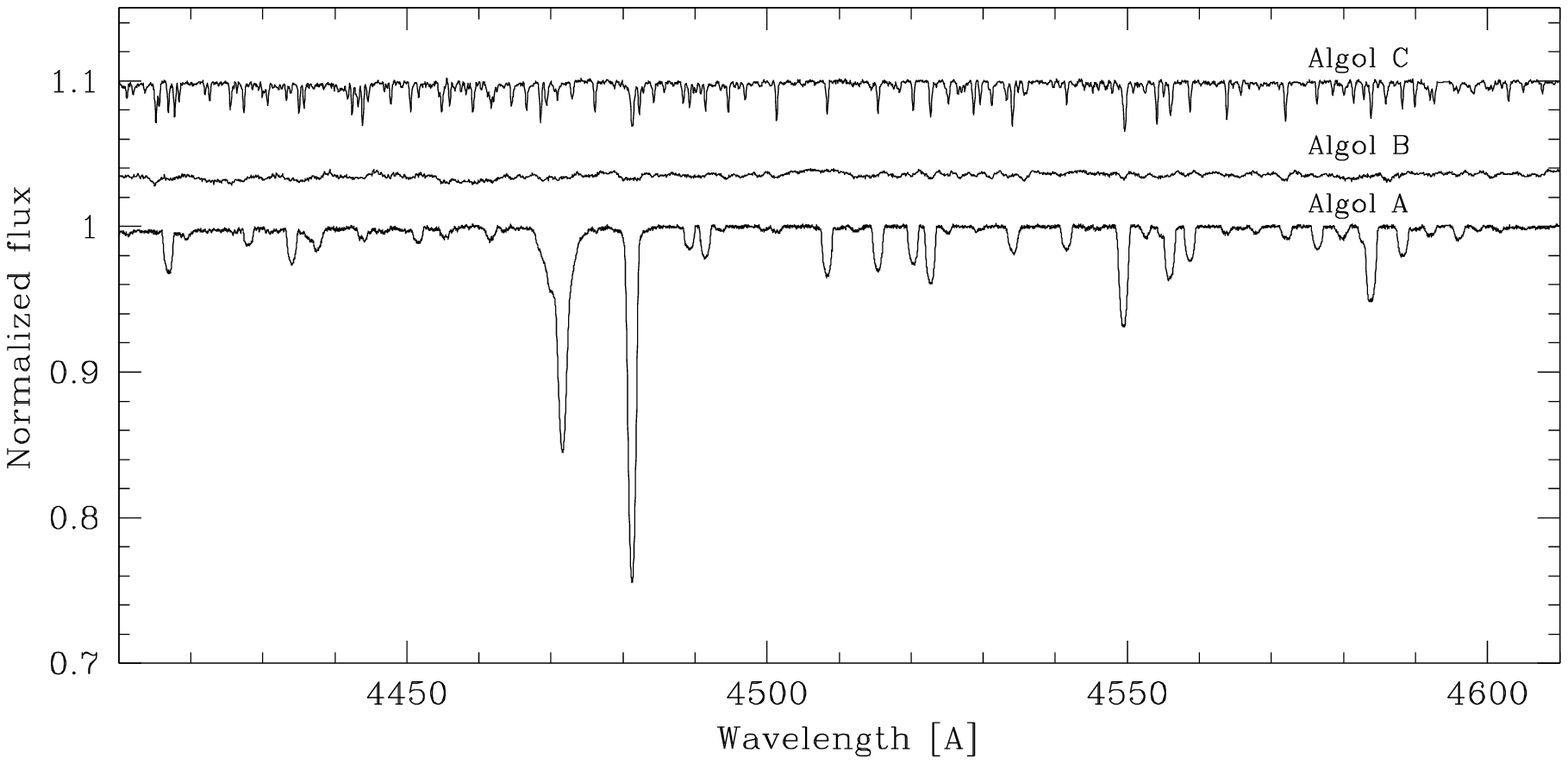}
\includegraphics[width=18cm]{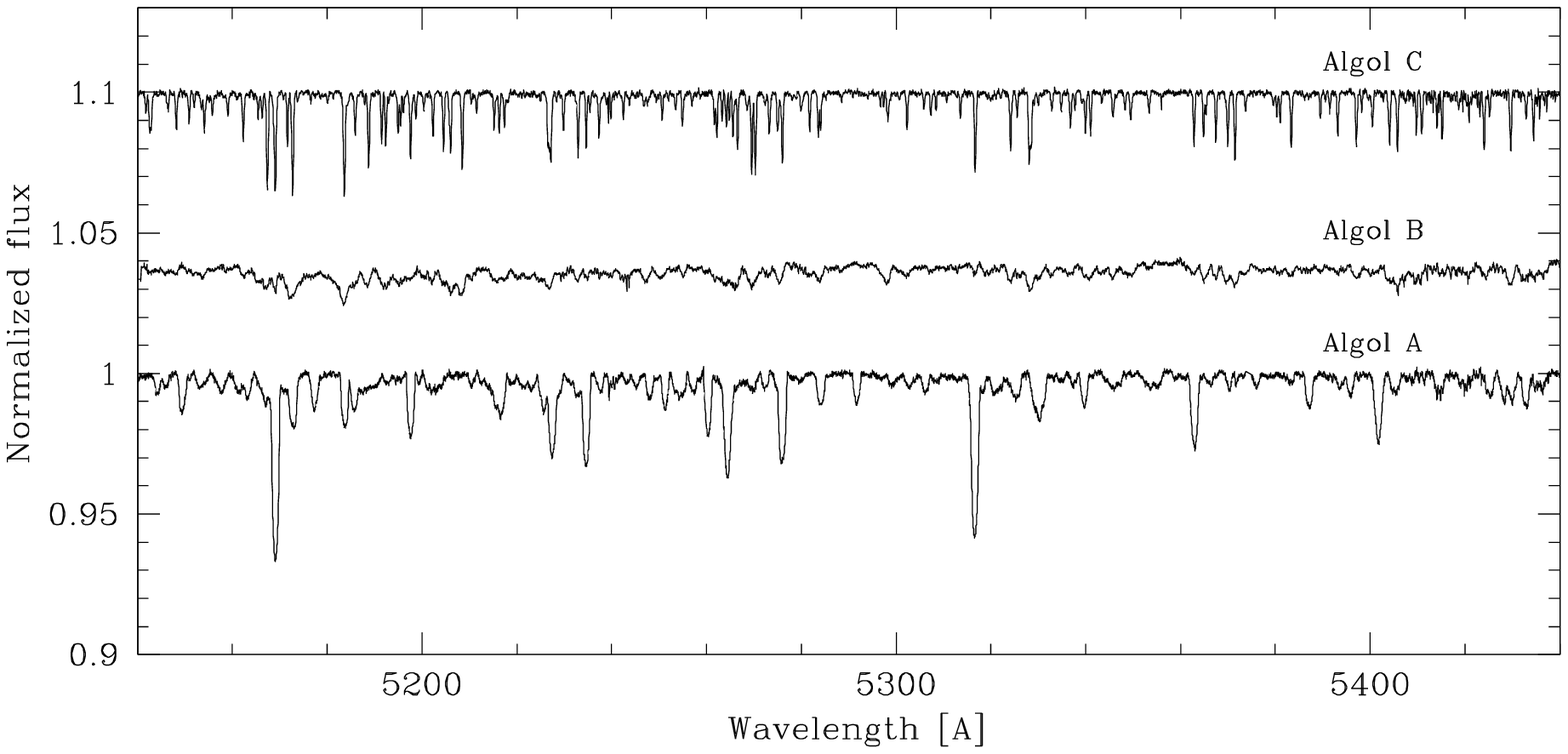}
\caption{Two portions of disentangled spectra for the three components of the Algol
system. The plots are not on the same scale, and disentangled spectra for Algol B and
C are shifted up by 0.04 and 0.10 of the continuum flux, respectively, for clarity.}
\end{figure*}

%%%%%%%%%%%%%%%%%%%%%%%%%%%%%%%%%%%%%%%%%%%%%%%%%%%%%%%%%%%%%%%%%%%%%%%%%%%%%%%%%%%%%%%%%%%%%%%%%%%%%%%%%%%%%%%%%%%%%%%%%%%%%%%%%%%%%%%%%%%%%%%%%%%%%

\section{Orbits and masses through spectral disentangling}            \label{sec:orbits}

The method of \spd\ enables determination of the orbital elements of binary and multiple 
systems, along with the simultaneous separation of the individual spectra of the components, 
in a self-consistent way (Simon \& Sturm 1994). In that sense it is a generalisation of the 
Doppler tomography method of Bagnuolo \& Gies (1991) which was the first successful 
reconstruction of individual spectra for binary stars but which relies on predetermined 
RVs for the components for each spectrum. In \spd\ only a time series of observed spectra 
are needed, with roughly uniform coverage of the RV motion of the stars (cf.\ Hensberge et al.\ 
2008). Optimisation of the RVs is bypassed in favour of directly fitting for the orbital 
elements, as first implemented by Simon \& Sturm (1994).

With no need for input template spectra, \spd\ does not suffer from biases due to template 
mismatch (cf.\ Hensberge \& Pavlovski 2007 and references therein). However, no comprehensive 
study of the error propagation in \spd\ and comparison 
to cross-correlation has yet been undertaken. Some initial studies have indicated the method 
is `well-behaved' (Hynes \& Maxted 1998; Iliji\'{c} et al.\ 2001; Hensberge \& Pavlovski 2007; 
Southworth \& Clausen 2007).

The \spd\ of Algol is challenging for the following reasons: (i) the system is triple; 
(ii) the secondary component is barely discernable in the optical spectra; (iii) the period 
of the outer orbit is long (680\,d); and (iv) eclipse spectra could not be used due to 
significant distortion of the line profiles  due to the Rossiter-McLaughlin effect 
(Rossiter 1924; McLaughlin 1924).

Spectra of several multiple systems have previously been successfully separated using 
different variants of the \spd\ method (cf.\ Fr\'{e}mat et al.\ 2005; Gonz\'{a}les et al.\ 
2006; Lee et al.\ 2008; Tamajo et al.\ 2012). Faint components with a fractional light 
contribution of only $\sim$5\% have previously been detected using \spd\ (e.g.\ Pavlovski 
et al.\ 2009; Lehmann et al.\ 2013; Tkachenko et al.\ 2014; Borkovits et al.\ 2014), 
including cases with only $\sim$2.5\% (Torres et al.\ 2014), and 1.5--2\% (Holmgren et 
al.\ 1999; Mayer et al.\ 2013) in the $V$ spectral region. As was shown in our preliminary 
reports on \spd\ of Algol (Pavlovski et al.\ 2010; Kolbas et al.\ 2012) we are pushing 
the limit of the method because of the extreme faintness of Algol B (see Sect.\ 5.1).

 The rotational distortion of the spectral lines during eclipse (the Rossiter-McLaughlin 
effect) violates the principal assumption of spectral disentangling that spectral line profiles 
are not intrinsically variable, so spectra obtained during eclipse cannot be used. 
Without such spectra there is insufficient variation in the light contributions of 
the components during the orbit, so the zeroth mode in the Fourier expansion is undetermined. 
This has the consequence that spectra can be only separated, and remain in the common continuum 
of the system. Renormalisation of the separated spectra of the individual components is discussed 
further in Sect.\ 5.1.

To perform \spd\ we used the code {\sc fdbinary}\footnote{{\tt hhtp://sail.zpf.fer.hr/fdbinary}} 
(Iliji\'{c} et al.\ 2004). {\sc fdbinary} implements disentangling in the Fourier domain 
(Hadrava 1995) using fast fourier transforms, making it computationally efficient. 
It can also account for two orbits and three individual spectra, as is required for Algol. 
Five spectral segments were selected, paying particular attention to including enough 
spectral lines of Algol B to allow its velocity semiamplitude to be precisely determined. 
For the blue spectral region this is challenging since the fractional light contribution 
is less than 1\% (see Sect.\ 5.1 and Fig.\ 2). The following spectral segments were used: 
4176.9--4277.0, 4400.2--4614.7, 4929.1--5118.9, 5140.6--5305.1 and 5311.0--5520.0 \AA. 
Only spectra taken outside eclipse were used, giving a total of 112 input spectra. 
To limit the computational demand we also stacked spectra taken in close succession, 
resulting in a total of 49 input spectra.  Such spectra typically covered a time interval 
of 3\,min, for which the RV of Algol B changes by a maximum of 0.5\kms, so velocity smearing 
is not a problem. The FIES and BOES datasets were also analysed first separately with the 
outer orbit fixed according to recent interferometric results (Czismadia et al.\ 2009; 
Zavala et al.\ 2010; Baron et al.\ 2012). The solution for the inner orbit was then used 
as an initial set of the orbital elements for solving both orbits simultaneously using 
the 49 spectra.

Hill et al.\ (1971) found that the inner orbit is slightly eccentric, $e_{\rm A-B} = 0.015\pm0.008$. 
In our \spd\ calculations the eccentricity of the inner orbit always converged to $e_{\rm A-B} 
= 0.0$ with a high confidence. Therefore, in the final set of the calculations, the eccentricity 
of the inner orbit was fixed to zero.

\begin{table}
\caption{The final solution for the orbital elements of the inner and outer
orbit in the Algol triple system as obtained by \spd. The periods are taken
from Baron et al.\ (2012), as well as the eccentricity of the outer orbit.
Trial calculations for the inner orbit has shown it is circular, and in the
final calculations $e_{\rm A-B}$ and $\omega_{A-B}$ were set to 0.}
\begin{tabular}{lcccc}
\hline
Quantity             & Notation            & Unit & Value    & Error \\
\hline
{\sc Inner orbit}    &                     &      &          &       \\
Period               & $P_{\rm A-B}$       & d    & 2.871362 & fixed \\
Eccentricity         & $e_{\rm A-B}$       &      & 0.       & fixed \\
Periastron longitude & $\omega_{\rm A-B}$  & deg  & 90.      & fixed \\
RV semiamplitude     & $K_{\rm A}$         & \kms & 44.1     & 0.2   \\
RV semiamplitude     & $K_{\rm B}$         & \kms & 194.2    & 1.2   \\
Mass ratio           & $q_{\rm A-B}$       &      & 0.227    & 0.005 \\
\hline
{\sc Outer orbit}    &                     &      &          &       \\
Period               & $P_{\rm AB-C}$      & d    & 680.168  & fixed \\
Time of periastron   & $T_{\rm AB-C, 0}$   & d    & 2454433.2& 1.1  \\
Eccentricity         & $e_{\rm AB-C}$      &      & 0.227    & fixed \\
Periastron longitude & $\omega_{\rm AB-C}$ & deg  & 138.1    & 0.6   \\
RV semiamplitude     & $K_{\rm AB}$        & \kms & 11.9     & 0.4   \\
RV semiamplitude     & $K_{\rm C}$         & \kms & 32.9     & 0.8   \\
Mass ratio           & $q_{\rm AB-C}$      &      & 0.364    & 0.009 \\
\hline
\end{tabular}
\end{table}

 Whilst the coverage of the orbital cycle for the inner orbit is good (Fig.~1), the phase 
distribution of the observations for the outer orbit is not sufficient to determine all 
parameters simultaneously, so we fixed the eccentricity of this orbit. Recent independent 
interferometric studies give a very consistent result on its value of $e_{\rm AB-C} = 
0.227\pm0.002$ (Baron et al.\ 2012). Since the {\sc simplex} algorithm is used for 
minimisation in {\sc fdbinary}, care must be taken to avoid becoming trapped in local 
minima. We used 50 runs with 1000 iterations each. Fixing $e_{\rm AB-C}$ made the convergence 
stable and consistent. Our final solution for the two orbits is given in Table~1. It represents 
the mean values from the solutions of \spd\ in the five selected spectral segments specified 
above. The errors quoted in Table~1 are standard deviations of the mean calculated from the 
solutions obtained for the five segments. More sophisticated error calculations for \spd, 
such as the jackknife method (Pavlovski et al.\ 2011, 2014), were beyond our 
 computational resources.

\begin{figure*}
\includegraphics[width=5cm]{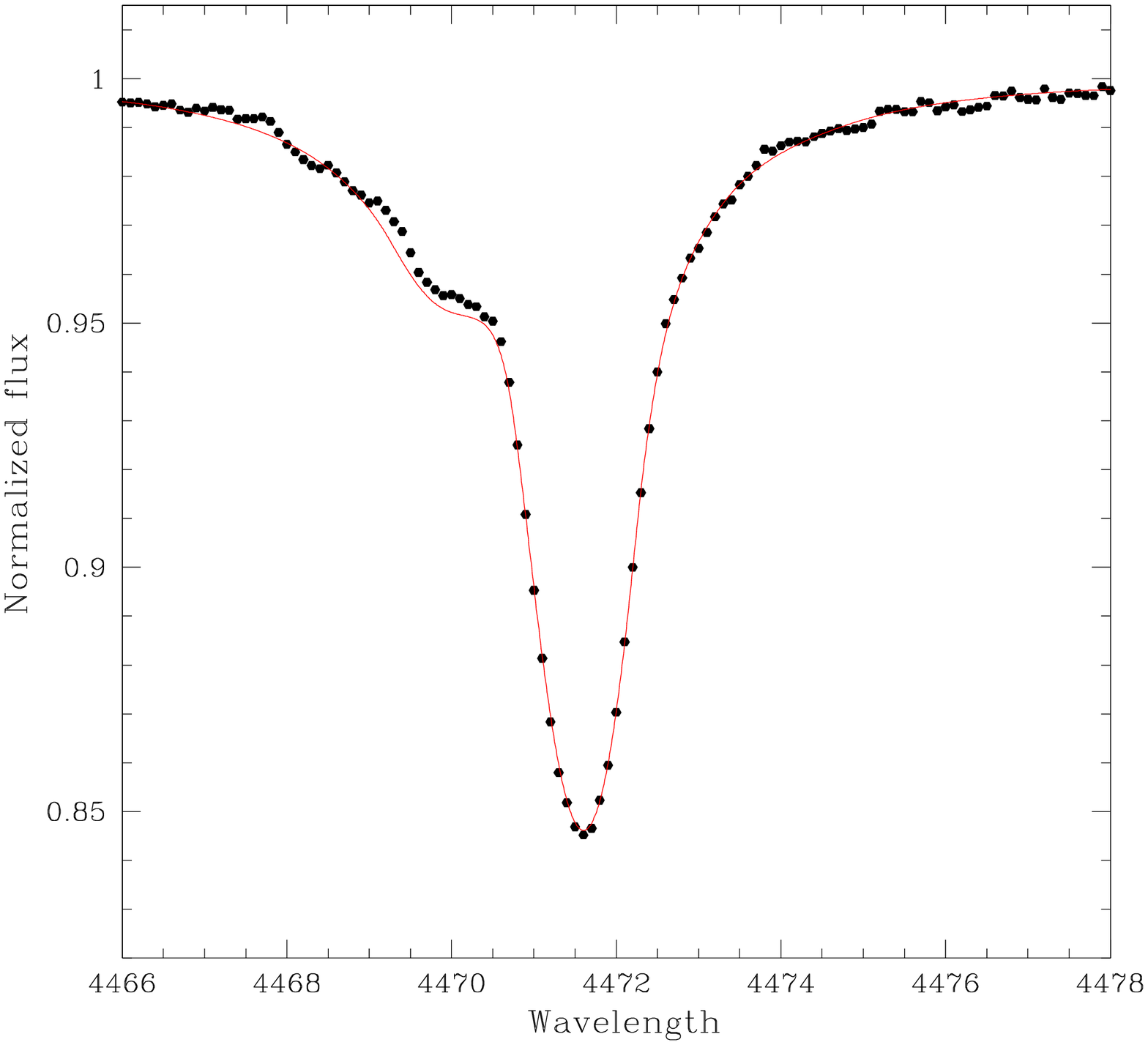}
\includegraphics[width=5cm]{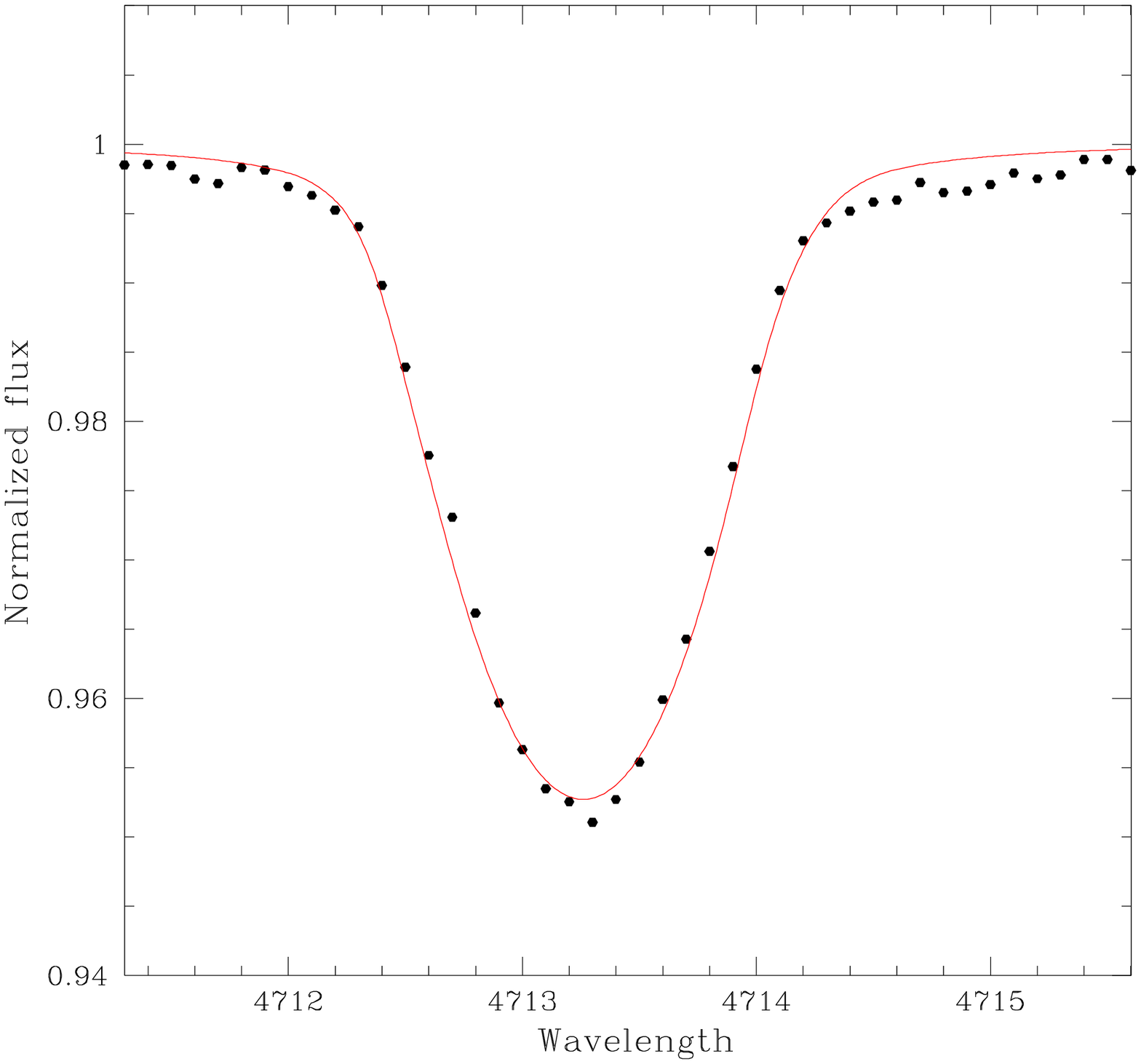}
\includegraphics[width=5cm]{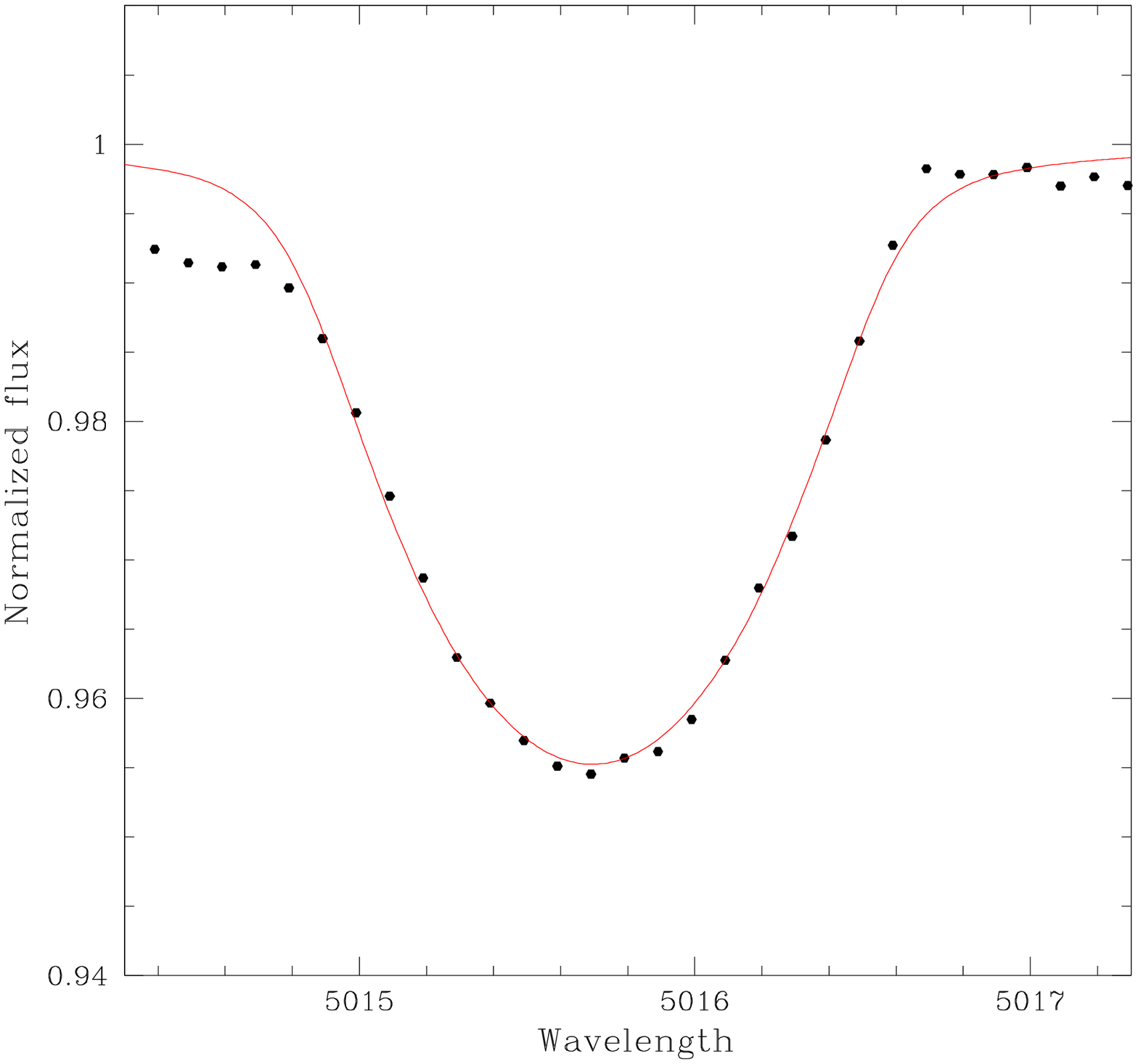}
\caption{Determination of the fractional light contribution for Algol A from optimal fits
of \ion{He}{i} lines in its disentangled spectrum. Synthetic NLTE \ion{He}{i} line profiles
were calculated for $\Teff = 12\,550$\,K, $\logg = 4.05$ and $v \sin i = 50.5$\kms.}
\end{figure*}

The comparison between the results presented in Table~1 and those from previous studies is 
interesting. First, we did not confirm the eccentricity for the inner orbit ($e_{\rm A-B}$) 
which was reported in some previous studies, but rejected in others. In an extensive spectroscopic 
study Hill et al.\ (1971) examined a series of medium-resolution photographic spectra obtained 
at the Dominion Astrophysical Observatory. They found $e_{\rm A-B} = 0.015\pm0.008$ but were 
not able to detect the spectrum of the secondary star. It is not clear if and how this might 
affect the determination of RVs for the primary. Also, it should be noted that Hill et al.\ 
(1971) attributed a 32\,year periodicity to apsidal motion in the inner orbit. The possibility 
of orbital eccentricity for the eclipsing pair has not been supported by past photometry of 
Algol (Wilson et al.\ 1972; Guinan et al.\ 1976; S\"{o}derhjelm 1980; Kim 1989).

The velocity semi-amplitude for Algol A found by Hill et al.\ (1971), $K_{\rm A} = 44.0\pm0.4$\kms,
 agrees within the uncertainties with the value derived in our work, $44.1\pm0.2$\kms. Also, it is 
encouraging that their velocity semi-amplitudes of the outer orbit, $K_{\rm AB} = 12.0\pm0.4$\kms\ 
and $K_{\rm C} = 31.6\pm1.2$\kms, are in agreement with our values, $K_{\rm AB} = 11.9\pm0.4$\kms\ 
and $K_{\rm C} = 32.9\pm0.8$\kms. Hill et al.\ (1971) were also able to determine the eccentricity 
of the outer orbit, $e_{\rm AB-C} = 0.23\pm0.04$, which was corroborated and improved by 
interferometric measurements (Zavala et al.\ 2010; Baron et al.\ 2012).

Two portions of disentangled spectra of the three components are presented in Fig.\ 2. The 
secondary spectrum is clearly isolated in the whole optical range, even though it contributes 
only 1\% of the light around 4500\,\AA\ and somewhat less than 2\% around 5500\,\AA\ (Sect.\ 5.1). 
The RV semi-amplitude for Algol B, $K_{\rm B}$, has two almost equally deep local minima around 
194 and 211\kms. It is difficult to trace the reasons for this ambiguity but, besides the small 
light contribution, these might include imperfect phase distribution of the observations for the 
inner orbit, gaps in the phase distribution for the outer orbit, and possible systematics due to 
the use of two spectrographs and varying observing conditions. The $\chi^{2}$ for the lower 
solution is slightly better, so we adopted this as our final solution, $K_{\rm B} = 194.2\pm1.2$\kms. 
This value agrees with the only previous measurement, $K_{\rm B} = 201\pm6$\kms\ 
(Tomkin \& Lambert 1978).

Tomkin \& Lambert (1978) calculated the spectroscopic mass ratio using the $K_{\rm A}$ determined 
by Hill et al.\ (1971) and their own $K_{\rm B}$ measurement, finding $q_{\rm A-B} = 0.219\pm0.007$. 
This mass ratio was subsequently used in many studies on Algol, as it was the first mass ratio 
found via dynamical effects. Our value is slightly higher, $q_{\rm A-B} = 0.227\pm0.002$. The 
mass ratio can be estimated from light curve for semi-detached eclipsing binaries, and there have 
been attempts to derive it this way. Kim (1989) used the `$q$-search method' to obtain exactly to 
the same value from $BV$ light curves, $q_{\rm phot} = 0.227\pm0.002$. Determination of the 
astrometric orbit from the interferometric measurements also can yield the mass ratio. 
Baron et al.\ (2012) solved the orbital elements for both orbits and the mass ratio of 
the inner orbit from the relative positions of the components derived from the visibility 
functions. In addition, they derived the mass ratio of the outer orbit independently from 
parallax assumptions. They found $q_{\rm A-B} = M_{\rm B}/M_{\rm A} = 0.219\pm0.017$ and 
$q_{\rm AB-C} = M_{\rm C}/(M_{\rm A}+M_{\rm B}) = 0.456\pm0.022$. This $q_{\rm A-B}$ agrees 
with our value to within its relatively large uncertainties.

The mass ratio for the outer orbit derived from these different techniques is highly discrepant 
and needs clarification. However, it is encouraging that spectroscopic analyses are consistent, 
as Hill et al.\ (1971) found $q_{\rm AB-C} = 0.380\pm0.051$, compared to our value of 
$q_{\rm AB-C} = 0.362\pm0.042$.

%%%%\section{Combined solution of astrometry and spectral disentangling}

%%%%%%%%%%%%%%%%%%%%%%%%%%%%%%%%%%%%%%%%%%%%%%%%%%%%%%%%%%%%%%%%%%%%%%%%%%%%%%%%%%%%%%%%%%%%%%%%%%%%%%%%%%%%%%%%%%%%%%%%%%%%%%%%%%%%%%%%%%%%%%%%%%%%%

\section{Atmospheric diagnostics}

\subsection{Renormalisation of disentangled spectra}

 Without eclipse spectra or significant variations in the fractional light contribution between 
the components, \spd\ can only be performed in the separation mode. The renormalisation of 
the individual spectra of the components can then be done using external information on the 
individual components' fractional contribution to the total light of the system.

A simple way is to use the light ratio between the components from the light curve solution 
for eclipsing binaries (cf.\ Hensberge et al.\ 2000). However, the light curve solution of 
Algol is rather uncertain because the eclipses are partial and there is `third light' from 
component C. On top of these obstacles, it is extremely difficult to acquire photometry of 
Algol due to its brightness and lack of suitable comparison stars within 10$^\circ$.

It is therefore not surprising that only a few complete light curves for Algol have been 
published. The most recent photometry (Kim 1989) was a series of photoelectric measurements 
in the $B$ and $V$ filters, obtained from 1982 to 1984 at Yonsei Observatory, Korea. The other 
light curves of Algol from the era of photoelectric photometry are those of Guinan et al.\ 
(1989) and Wilson et al.\ (1972). In addition, Stebbins secured an incomplete light curve in 
multiple passbands from 1949 to 1951 (Stebbins \& Gordon 1975), Al-Naimiy et al.\ (1985) 
obtained unpublished observations in 1981, and light curves in the ultraviolet (Eaton 1975), 
and near-infrared (Chen \& Reuning 1966) exist.

\begin{table} \begin{center}
\caption{Summary of the fractional light contribution of the components of Algol to the
total light of the system determined by different techniques. `Photometry' indicates
light curve analysis. Only $B$ and $V$ photometry are listed as this is the wavelength
region covered by our spectra.
References: (1) Wilson et al.\ (1972); (2) Demircan (1978); (3) Richards et al.\ (1988);
(4) Kim (1989); (5) Zavala et al.\ (2010); (6) This work.}
\begin{tabular}{lccc} \hline
Method            &  $B$     &     $V$     &    Source \\
\hline
Photometry        & 0.940$\;$0.010$\;$0.050  & 0.893$\;$0.029$\;$0.078 &  (1)   \\
Photometry        & 0.909$\;$0.041$\;$0.050  & 0.857$\;$0.065$\;$0.078 &  (2)   \\
Photometry        & 0.929$\;$0.012$\;$0.059  & 0.894$\;$0.035$\;$0.071 &  (3)   \\
Photometry        & 0.854$\;$0.040$\;$0.106  & 0.814$\;$0.067$\;$0.119 &  (4)   \\
Interferometry    &                          & 0.863$\;$0.063$\;$0.065 &  (5)   \\
Spectroscopy      & 0.943$\;$0.008$\;$0.049  & 0.915$\;$0.018$\;$0.067 &  (6)   \\
\hline \end{tabular} \end{center} \end{table}

Previous light curve analyses have not yet given a unique solution for Algol, primarily because 
of its complexity. Solutions of partially eclipsing binary stars inevitably suffer from degeneracy, 
which should be lifted using additional information (Garcia et al.\ 2014; Southworth et al.\ 2007). 
This is illustrated in Table~2 in which fractional light contributions (light dilution factors) 
for Algol are listed from several works. Wilson et al.\ (1972) used an estimate of the third 
light contribution derived spectroscopically by Fletcher (1964). The same approach was followed 
by Demircan (1978) who analysed the photometry published in Wilson et al.\ (1972). Since the 
secondary component had not yet been detected in the spectrum of Algol, the mass ratio between 
the components of the inner eclipsing pair had to be assumed from other considerations. But it 
is evident from the solutions derived in Richards et al.\ (1988) and Kim (1989) that the 
dynamical mass ratio measured by Tomkin \& Lambert (1978) did not help to provide a consistent 
description of the Algol system. Richards et al.\ (1988) calculated a grid of models with 
different assumptions for the radiative properties of the third star, whilst Kim (1989) 
attempted to derive its contribution directly from his $BV$ light curves. In this context 
we also list estimates of the light ratios measured in the interferometric observations 
by Zavala et al.\ (2010). Their estimates suffered from large uncertainties, and the 
fractional light contribution of $\sim$6\% for Algol B is certainly too large.

Facing all these uncertainties in the determination of the fractional light contributions 
of the components, we rely on the extraction of these quantities from the disentangled 
spectra themselves. The information on the light dilution factor is contained in the spectral 
line depths in the disentangled (separated) spectra of an individual component. But caution is 
needed at this point since other effects could change the line depth in stellar spectrum and 
mimic light dilution effects in binary or multiple systems, such as metal abundance or chemical
 peculiarity of the component(s). Algol C is often characterised as an Am star, beginning with 
the spectroscopic analysis by Fletcher (1964).

In the optical spectrum of Algol the most prominent lines are \ion{H}{i} and \ion{He}{i} 
originating in the photosphere of the primary component (cf.\ Struve \& Sahade 1957). Since 
\ion{He}{i} lines are present in only the primary star's spectrum, we used them for the 
determination of its effective temperature, $T_{\rm eff, A}$, and fractional light contribution,
 $lf_{\rm A}$. We used the program {\sc starfit} (Kolbas et al.\ 2014) which compares 
disentangled component spectra to a grid of calculated theoretical spectra. {\sc starfit} can 
handle the following parameters: effective temperature $T_{\rm eff}$, surface gravity $\logg$, 
fractional light contribution $lf$, projected rotational velocity $v\sin i$, relative velocity 
shift between disentangled spectrum and rest-frame laboratory wavelengths, $v_{\rm 0}$, and 
continuum corrections, $cc$. Optimisation is performed by a genetic algorithm (Charbonneau 1995), 
and can be done in constrained mode with disentangled spectra of the components for which the 
sum of the fractional light contributions should be $\leq 1$ (Tamajo et al.\ 2011). Grids of 
theoretical spectra were calculated in LTE (local thermodynamic equilibrium) using the 
{\sc uclsyn} code (Smith 1992, Smalley et al.\ 2001)

$T_{\rm eff, A}$ is constrained to be in the range 12\,000 to 13\,000 K (Richards et al.\ 1988; 
Kim 1989). Our disentangled spectra contain three \ion{He}{i} lines suitable for the optimisation, 
at 4471, 4713 and 5015\,\AA. Unfortunately, the \ion{He}{i} lines at 4388, 4920, 5047, and 
5788\,\AA\ are either on the wings of Balmer lines, or are contaminated by metal or telluric 
lines, so are not suitable. To facilitate optimisation we first determined $v\sin i$ by optimal
 fitting of unblended metal lines in the primary's disentangled spectrum ($v\sin i$ or {\sc fwhm} 
does not depend on the light dilution). In hot stars, a degeneracy exists between \Teff\ and 
\logg\ for \ion{H}{i} and \ion{He}{i} lines, so we fixed \logg\ in our calculations to 
$\log g_{\rm A}  = 4.05$ (Richards et al.\ 1988). Also, it should be noted that model 
atmospheres are calculated assuming the `standard' (solar) helium abundance with the 
fraction of helium atoms $N_{\rm He}/(N_{\rm H} + N_{\rm He}) = 0.089$.

We found that the three \ion{He}{i} line profiles are not well and consistently reproduced in 
the LTE approximation.  This is not a surprise because discrepancies in the \ion{He}{i} line 
profiles are the most severe at the low and high end of the temperature sequence (Auer \& Mihalas 
1970, 1973). Thus we decided to calculate \ion{He}{i} line profiles in non-LTE. Our grid of 
theoretical spectra (Kolbas et al.\ 2014) was extended down to $T_{\rm eff} = 12\,000$ K. 
This is below the usual borderline used for NLTE spectrum synthesis of $T_{\rm eff} = 15\,000$ K. 
Almost perfect fits were achieved with \ion{He}{i} line profiles in NLTE (Fig.~3, solid lines). 
Optimisation was performed separately for each line to find the wavelength-dependent fractional 
light contribution. An excellent convergence was found for almost same effective temperature, 
$T_{\rm eff, A} = 12\,600\pm90$\,K, $lf(4471+4713) = 0.943\pm0.002$, and $lf(5015) = 
0.915\pm0.002$. Uncertanties in \Teff\ and $lf$ were calculated with a Markov chain Monte 
Carlo technique (Ivezi\'{c} et al.\ 2014). A refined \Teff\ measurement for Algol A is discussed 
below (Sect.\ 5.2).

As described in Sect.~4, \spd\ enabled us to reconstruct almost the entire optical spectrum of 
Algol B, the first time this has been achieved (Figs.\ 2 and 4). A first attempt to match it to 
theoretical spectra with $T_{\rm eff, B} \sim 4500$\,K indicated a low fractional light 
contribution, notably less than indicated by light curve solutions (Table~2). Again, the 
program {\sc starfit} was used with only surface gravity as a fixed parameter, $\log g_{\rm B} 
= 3.11$ (Richards et al.\ 1988). The projected rotational velocity converged to $v_{\rm B} 
\sin i_{\rm B} = 62\pm2$\kms, in good agreement with the expected synchronous rotational 
velocity for this star ($v_{\rm synch} = 61.8\pm0.5$\kms. The optimal fractional light 
contributions of Algol B are: $ldf(4500) = 0.008\pm0.001$ and $ldf(5500) = 0.018\pm0.001$, 
with $T_{\rm eff, B} = 4900\pm300$ K. This is the most unexpected result of our study but 
quite secure since the spectral lines for the late G or early K subgiant are intrinsically 
deep and, for the expected $v \sin i$, the light dilution is certainly no more than 2\% in 
$V$ and less than 1\% in $B$.

\begin{figure}
\includegraphics[width=8.5cm]{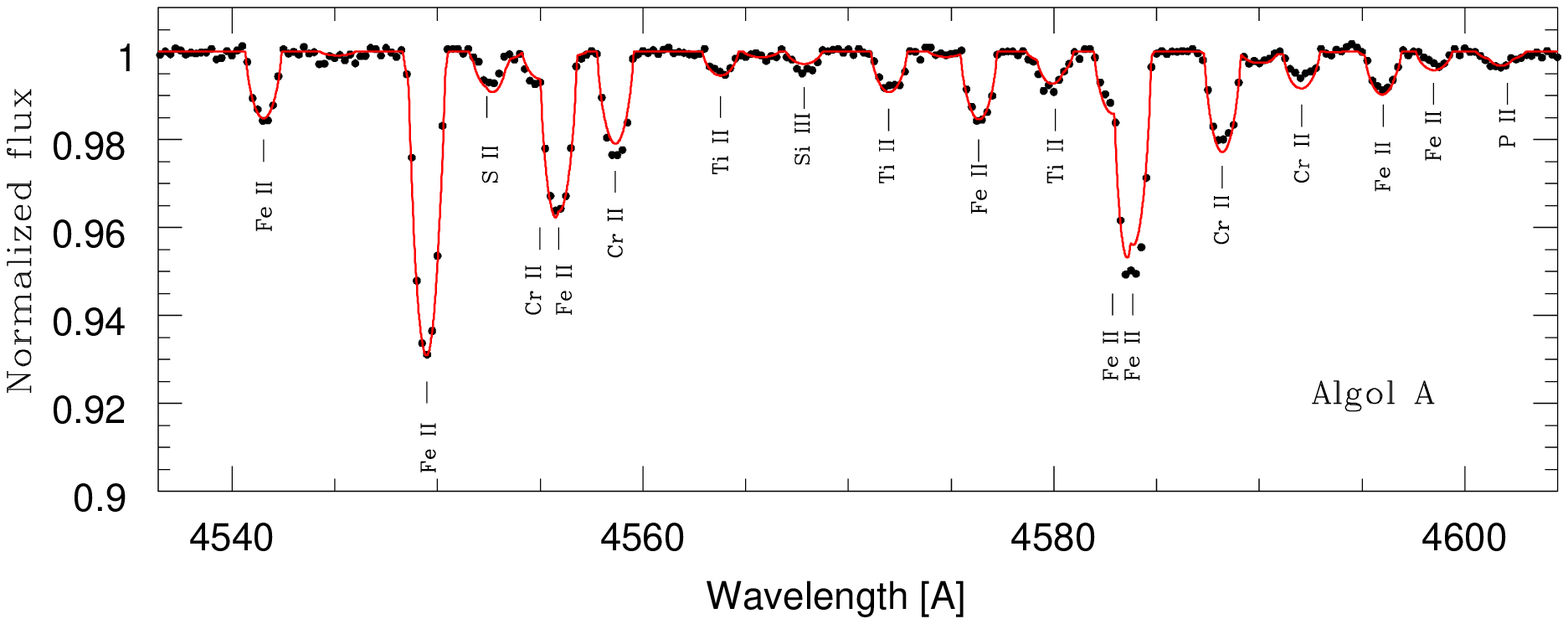}
\includegraphics[width=8.5cm]{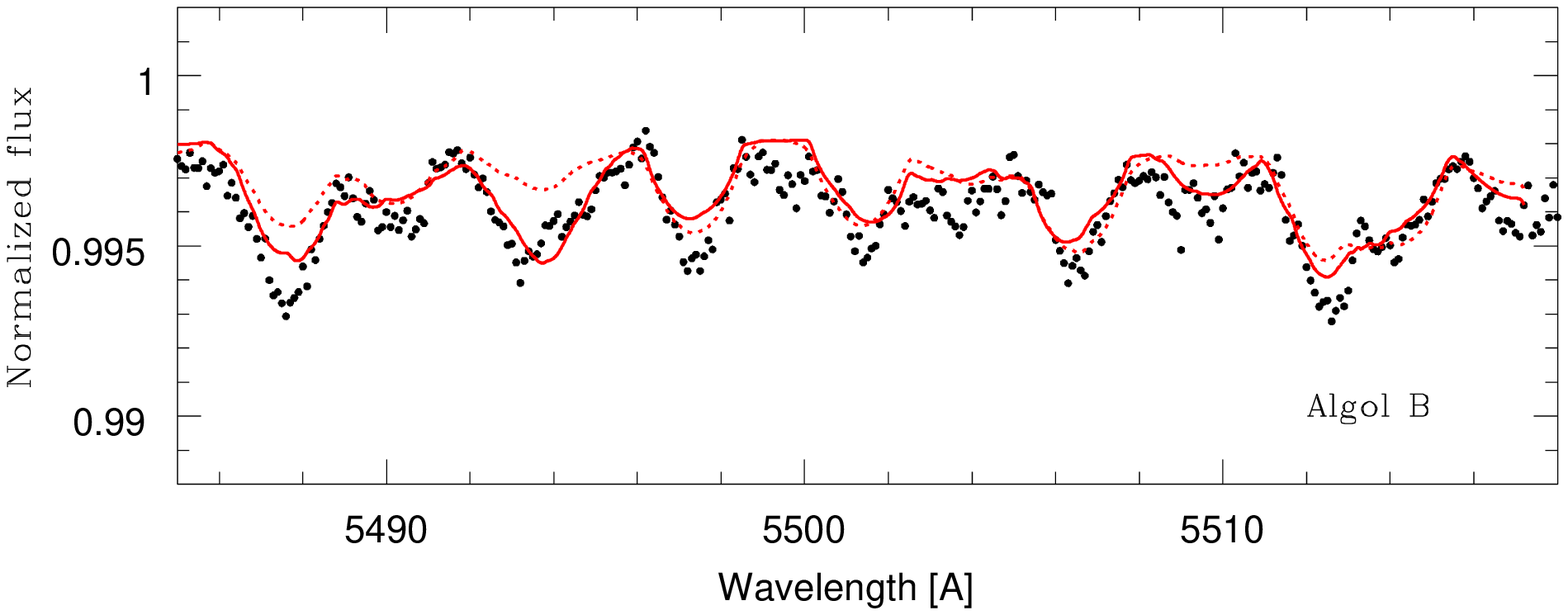}                     %%%% {newsecfit2x.ps}
\includegraphics[width=8.5cm]{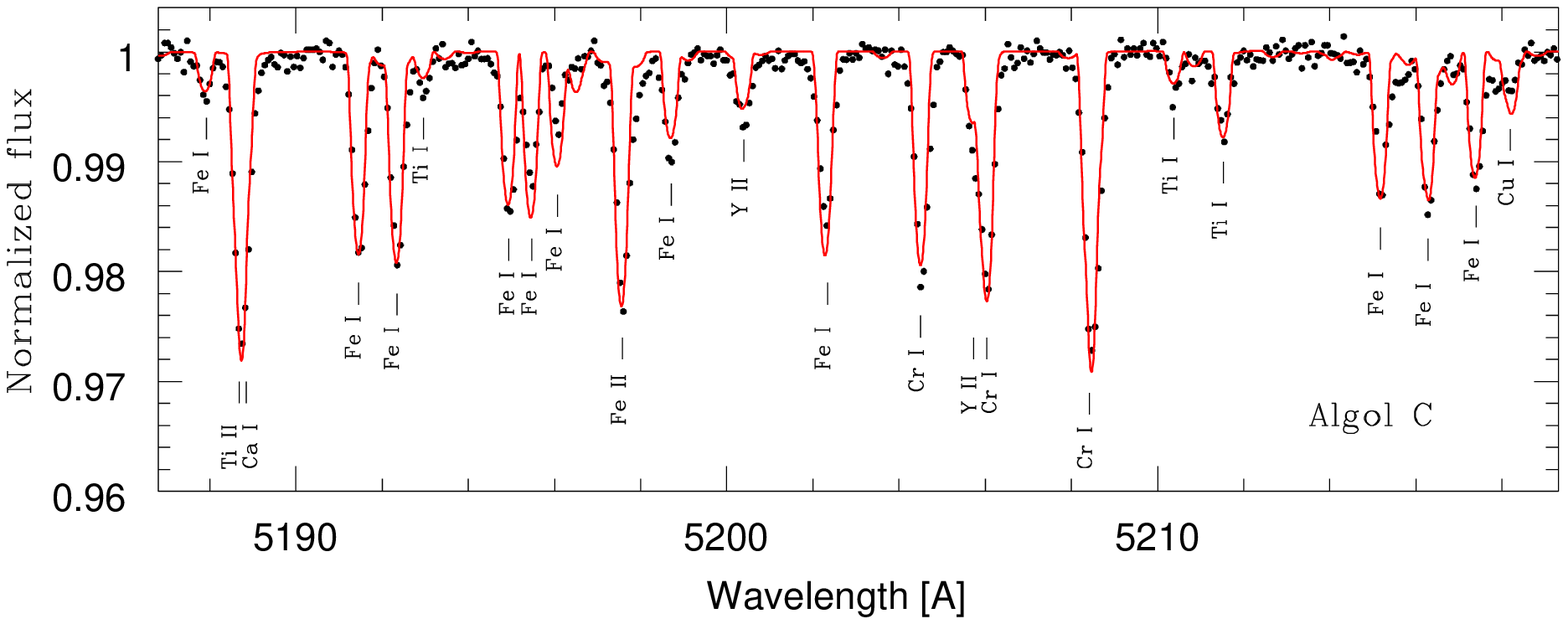}
\caption{Optimal fits of the disentangled spectra of the components in Algol.
From top to bottom are shown Algol A, Algol B, and Algol C, respectively.
These short portions of the components' disentangled spectra show the quality
of the optimal fitting for the atmospheric parameters listed in Table 5,
except for Algol B (middle panel) for which best fits for two \Teff s are
shown (solid line 5000\,K, and dotted line  4000\,K).}
\end{figure}

\subsection{Effective temperatures for the components}

Detailed spectral analysis is possible for Algol A and Algol C, making it possible to fine-tune 
their \Teff\ measurements. The values determined in Sect.~5.1 served as starting points. 
A spectral analysis of the entire disentangled spectral range ($4400$--$5800$\,\AA) was 
performed, using {\sc uclsyn}. The equivalent widths (EWs) were measured for suitable lines 
in the renormalised disentangled spectra. Model atmospheres were calculated in LTE with the 
program {\sc atlas9} (Kurucz 1979), and abundances calculated for the measured EWs. Since Fe 
lines are the most numerous in the spectra of both stars, they served for the determination 
of $T_{\rm eff}$ and microturbulent velocity, $\xi_{\rm t}$. The spectrum of Algol A contains 
many \ion{Fe}{ii} lines, but those of \ion{Fe}{i} are few and weak. That of Algol C contains 
many \ion{Fe}{i} lines but also a substantial number of \ion{Fe}{ii} lines, which means 
fine-tuning of $T_{\rm eff, C}$ is possible using the Fe ionisation balance  (see Fig.~4).

We determined $T_{\rm eff}$ and $\xi_{\rm t}$ in a few iteration steps. $T_{\rm eff, A}$ was 
tuned so there was no correlation between \ion{Fe}{ii} abundance and the excitation potential, 
EP, whilst the microturbulence, $\xi_{\rm t, A}$ was found by requiring the \ion{Fe}{ii} 
abundance to be independent of EW. We found $T_{\rm eff, A} = 12\,550\pm120$\,K and 
$\xi_{\rm t, A} = 0.4\pm0.2$\kms. The uncertainties were calculated from the uncertainties 
in the determination of the slopes of $\log \epsilon$(\ion{Fe}{ii}) versus EP for \Teff\ 
and $\log \epsilon$(\ion{Fe}{ii}) versus EW for $\xi_{\rm t}$.

The atmospheric parameters for Algol C were determined in the same manner. Moreover, the Fe 
ionisation balance could be used for fine-tuning $T_{\rm eff, C}$. Our final results are 
$T_{\rm eff, C} = 7\,540\pm80$\,K, and $\xi_{\rm t, C} = 1.64\pm0.08$\kms. The uncertainties 
were derived in the same way as for Algol A, except that \ion{Fe}{i} was used. The ionisation 
balance of Fe is well satisfied with a difference in abundance derived from \ion{Fe}{i} and 
\ion{Fe}{ii} lines of only $\Delta \log \epsilon ({\rm Fe}) = 0.05\pm0.11$.

Since we have renormalised the disentangled spectra of the components with the light dilution 
factors determined from these spectra themselves, the wavelength dependence of the Fe abundance 
could be an important check of the correctness of the procedure. \ion{Fe}{ii} and \ion{Fe}{i} 
lines are well distributed in the optical spectra of Algol A and Algol C, respectively. This 
test is more sensitive for Algol C as the multiplication factor needed to normalise its spectrum 
to unit continuum is much higher than in the case of Algol A, $\sim$16.5 versus $\sim$1.1. 
As is illustrated in Fig.~5 no wavelength dependence of iron abundance is present for either 
of the components. This provides encouraging support for the reliability of our estimates for 
the luminosity contribution of the components to the total light of the system. The $\xi_t$ 
for Algol A is also in agreement with that generally found for late B-type stars 
(Fossati et al.\ 2009), which is usually considerably less than 1\kms.

\begin{figure}
\includegraphics[width=8.5cm]{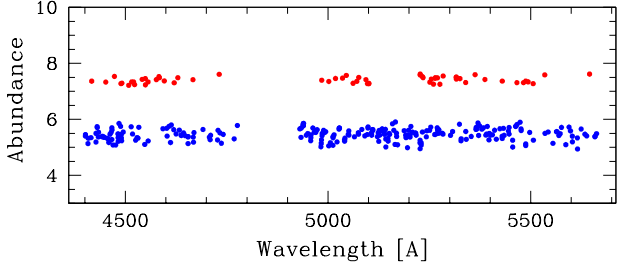} %% with error bars
\caption{Dependence of iron abundance on wavelength for Algol A (filled red circles) and
Algol C (filled blue circles). In the case of Algol A, the abundances of \ion{Fe}{ii} are
shown. For Algol C, \ion{Fe}{i} abundances are shown. The scatter is large for Algol C because
the S/N of the renormalised disentangled spectrum of this component is much less than that
for Algol A due to the differing fractional contributions to the total light of the system.}
\label{secfit}
\end{figure}

\subsection{The elemental composition and metallicity}

\begin{figure}
\includegraphics[width=8.5cm]{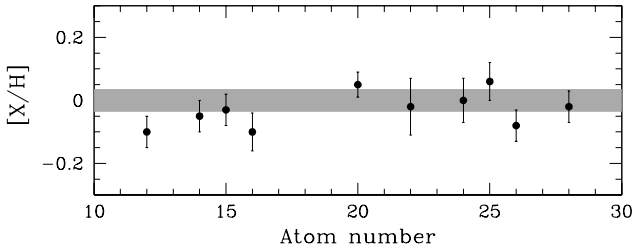} %% with error bars
\caption{Abundance pattern measured for the photospheric composition of Algol A (symbols)
compared to the standard solar composition (grey shading) which represents the 1$\sigma$
uncertainty in the solar composition determined by Asplund et al.\ (2009).}
\end{figure}

\begin{table} \centering
\caption{\label{tab:abund:a} Photospheric abundances derived for Algol A.
Abundances are expressed relative to the abundance of hydrogen (column 2),
$\log \epsilon(H) = 12.0$. The third column gives the number of lines used,
and abundances relative to the Sun are given in the fourth column.}
\begin{tabular}{lcrcc} \hline
El.\  & $\log \epsilon({\rm X})$ & $N$ & [X/H] &    \\
\hline
C   &  8.64$\pm$0.02  &  3  &  0.21$\pm$0.05  \\
N   &  7.92$\pm$0.04  &  4  &  0.09$\pm$0.06  \\
O   &  8.93$\pm$0.02  & 11  &  0.24$\pm$0.05  \\
Mg  &  7.49$\pm$0.03  &  7  & -0.10$\pm$0.05  \\
Si  &  7.46$\pm$0.04  & 14  & -0.05$\pm$0.05  \\
P   &  5.38$\pm$0.05  &  3  & -0.03$\pm$0.05  \\
S   &  7.02$\pm$0.02  & 21  & -0.10$\pm$0.06  \\
Ca  &  6.37$\pm$0.02  &  3  &  0.05$\pm$0.04  \\
Ti  &  4.91$\pm$0.03  & 11  & -0.02$\pm$0.09  \\
Cr  &  5.62$\pm$0.04  &  9  &  0.00$\pm$0.07  \\
Mn  &  5.48$\pm$0.05  &  5  &  0.06$\pm$0.06  \\
Fe  &  7.39$\pm$0.02  & 70  & -0.08$\pm$0.05  \\
Ni  &  6.18$\pm$0.02  & 17  & -0.02$\pm$0.05  \\
\hline \end{tabular} \end{table}

\begin{table}
\centering \caption{Photospheric abundances derived for Algol C.
Explanations are the same as for Table\,\ref{tab:abund:a}.}
\begin{tabular}{lcrcc} \hline
El.\  & $\log \epsilon({\rm X})$ & $N$ & [X/H] &    \\
\hline
C   &  8.53$\pm$0.04  &  21  &  0.10$\pm$0.06  \\
O   &  8.72$\pm$0.06  &   6  &  0.03$\pm$0.08  \\
Na  &  6.31$\pm$0.08  &   7  &  0.10$\pm$0.09  \\
Mg  &  7.50$\pm$0.09  &   7  & -0.09$\pm$0.10  \\
Si  &  7.58$\pm$0.06  &  23  &  0.07$\pm$0.07  \\
S   &  7.16$\pm$0.04  &  10  &  0.04$\pm$0.05  \\
Ca  &  6.37$\pm$0.09  &  21  &  0.02$\pm$0.09  \\
Sc  &  3.03$\pm$0.07  &  12  & -0.13$\pm$0.08  \\
Ti  &  5.01$\pm$0.07  & 106  &  0.08$\pm$0.08  \\
V   &  4.10$\pm$0.10  &  17  &  0.21$\pm$0.13  \\
Cr  &  5.69$\pm$0.07  &  97  &  0.07$\pm$0.08  \\
Mn  &  5.33$\pm$0.08  &  23  & -0.09$\pm$0.09  \\
Fe  &  7.45$\pm$0.08  & 296  & -0.02$\pm$0.09  \\
Co  &  5.03$\pm$0.07  &  21  &  0.10$\pm$0.09  \\
Ni  &  6.25$\pm$0.07  & 161  &  0.05$\pm$0.08  \\
Cu  &  4.19$\pm$0.13  &   3  &  0.01$\pm$0.14  \\
Y   &  2.32$\pm$0.06  &  14  &  0.11$\pm$0.08  \\
Zr  &  2.59$\pm$0.06  &  11  &  0.09$\pm$0.07  \\
Ba  &  2.38$\pm$0.34  &   3  &  0.29$\pm$0.18  \\
La  &  1.35$\pm$0.07  &   6  &  0.23$\pm$0.08  \\
Ce  &  1.69$\pm$0.06  &  16  &  0.11$\pm$0.07  \\
Nd  &  1.57$\pm$0.08  &   8  &  0.15$\pm$0.09  \\
\hline \end{tabular} \end{table}

\begin{figure}
\begin{tabular}{c}
\includegraphics[width=8.5cm]{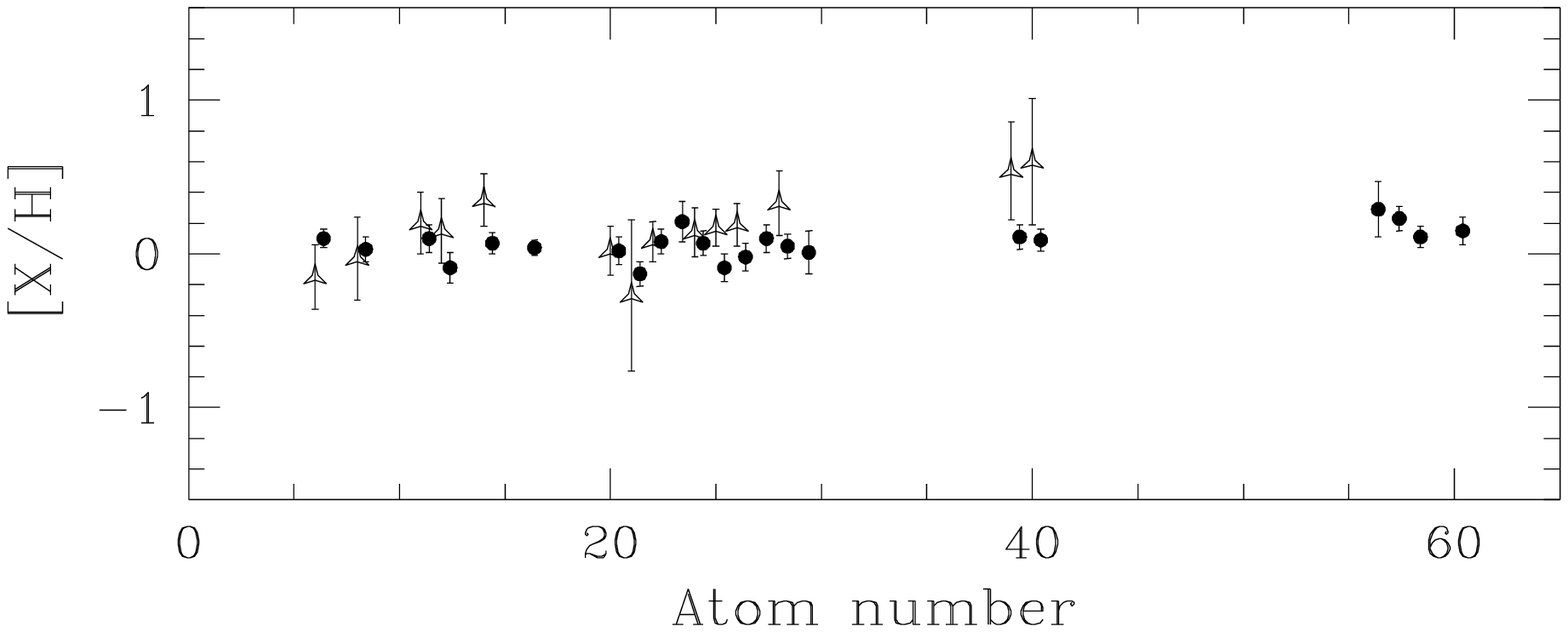} \\
\includegraphics[width=8.5cm]{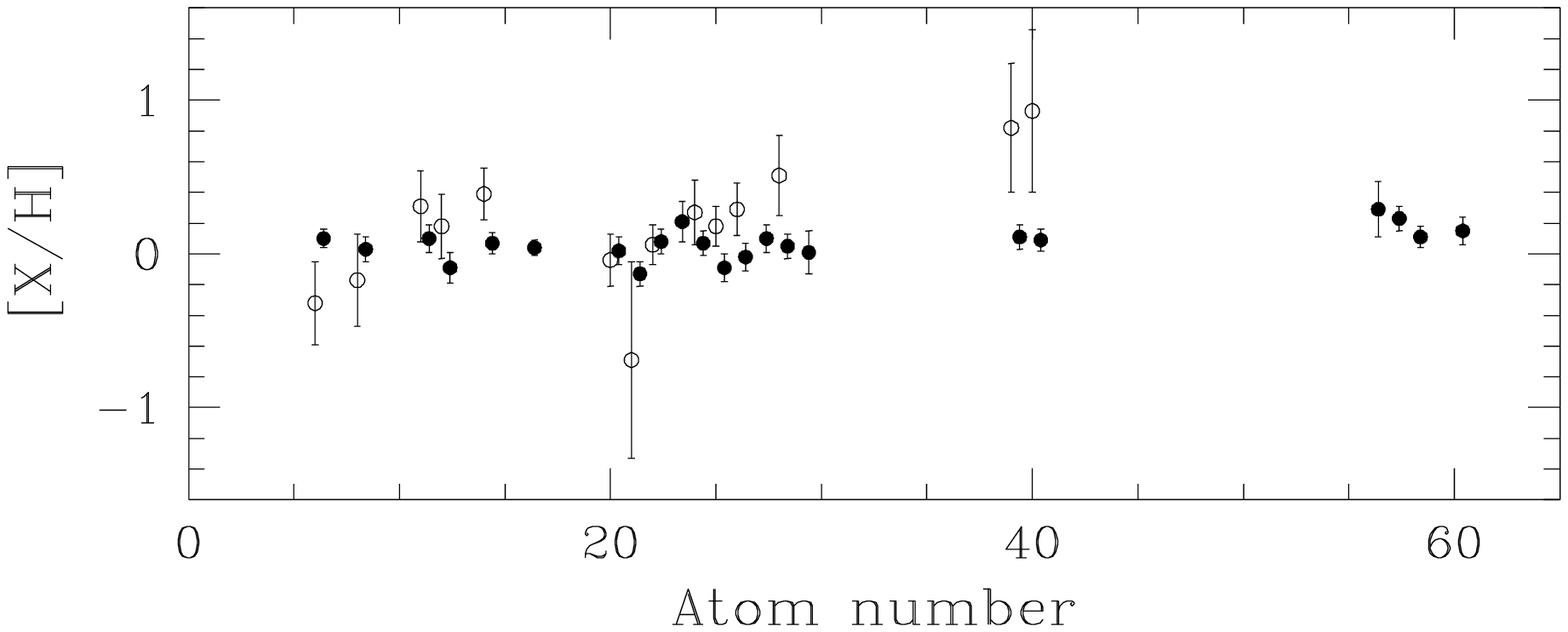}
\end{tabular}
\caption{Abundance pattern measured for the photospheric composition of
Algol C (black dots) compared to the average abundance pattern of `normal'
A-type stars (upper panel,  open triangles) and Am stars (lower panel,
 open squares), following Gebran et al.\ (2010).}
\end{figure}

The number of spectral lines available for the determination of the photospheric chemical 
composition of Algol A are rather limited. This does have the advantage that line blending 
and smearing do not prevent detailed analysis despite the relatively high $v\sin i$ of 50\kms. 
Beside the strong hydrogen (not disentangled) and helium lines, the most prominent lines are 
of \ion{Fe}{ii}. For the \Teff\ of Algol A, spectral lines of C, N and O are rather weak: 
the EWs of C lines are all $<$4.5\,m\AA\ except for the \ion{C}{ii} 4267\,\AA\ line, 
$<$3\,m\AA\ for N, and $<$15\,m\AA\ for O. Therefore, abundance estimates for these 
elements should be considered with caution, except for \ion{C}{ii} 4267\,\AA\ whose 
EW of 44.8\,m\AA\ gives a carbon abundance of $\log \epsilon({\rm C}) = 8.27\pm0.06$. 
This is $0.16\pm0.08$ dex less than solar (Asplund et al.\ 2009). A deficiency of carbon 
is also indicated from the comparison of the EW for the \ion{C}{ii} 4267\,\AA\ line to EWs 
of standards from Tomkin et al.\ (1993) and \.{I}bano\v{g}lu et al.\ (2012).

The mean photospheric metal abundance of Algol A has been calculated for all elements listed 
in Table~3 except CNO. The abundances of these elements were not used because of the possibility 
of changes caused by mass transfer, and the small EWs of the lines on which they are based. 
The remaining ten elements give an average abundance of $\mathrm{[M/H]_{\rm A}} = -0.03\pm0.08$. 
This is less than 1$\sigma$ away from the $\mathrm{[Fe/H]_{\rm A}} = -0.08\pm0.05$ value derived 
from the most numerous lines. We calculated the bulk metallicity using the approximate equation, 
$Z = Z_{\odot}\, 10^{-[{\rm M/H}]}$, finding it to be solar to within its uncertainty: 
$Z_{\rm A} = 0.014\pm0.002$ (using the present day solar metallicity $Z_{\odot} = 0.0134$ from 
Asplund  et al.\ 2009). The measured elemental abundances in the photosphere of Algol A are 
displayed in Fig.~6 relative to standard solar values (Asplund et al.\ 2009).

The rich and relatively unbroadened ($v_{\rm C} \sin i_{\rm C} \sim 12$\kms) spectrum of Algol C 
made it possible to determine the photospheric abundances for 20 elements. Seven appear in two 
ionisation stages (neutral and singly-ionised). In the spectral range studied (4500--5700\,\AA) 
the most numerous lines are of \ion{Fe}{i}, \ion{Ni}{i}, \ion{Cr}{i} and \ion{Ti}{ii}. The 
measured abundances are listed in Table~4, along with the number of spectral lines used, 
and abundances relative to standard solar (Asplund et al.\ 2009). The slightly larger 
uncertainties compared to Algol A, despite having more lines available, are due to the 
smaller S/N of the renormalised disentangled spectrum after multiplication by $\sim$16 
to put the continuum level to unity. In Fig.~7 we compare the photospheric composition 
of Algol C with the abundance pattern derived in Gebran et al.\ (2010) for a sample of 
`normal' A-type stars (upper panel) and Am stars (lower panel). The abundances in Algol C 
do not show the principal characteristics of the Am phenomenon, i.e.\ strong Sc underabundance, 
often a Ca underabundance, and a moderate to strong overabundance of iron-group elements. 
The mean metallicity for Algol C is consistent with solar, at $\mathrm{[M/H]_{\rm C}} = 
0.04\pm0.09$. The corresponding bulk metallicity is $Z_{\rm C} = 0.012\pm0.002$.

The photospheric elemental compositions of Algol A and Algol C are the same to within 
the 1$\sigma$ uncertainties, with a formal difference of $\Delta \mathrm{[M/H]} = 
\mathrm{[M/H]}_{\rm A} - \mathrm{[M/H]}_{\rm B} = -0.07\pm0.12$. Both correspond to 
the standard solar composition (Asplund et al.\ 2009).

\begin{table*}
\caption{Astrophysical quantities for the components of Algol triple system
derived in this work (masses, \Teff s, $v \sin i$ values, or calculated in
conjuction with previous studies (radii, \logg\ values, synchronous velocities).
The radius for Algol C is from the interferometric study by Baron et al.\ (2012).
All other quantites are determined or calculated in this work.}
\begin{tabular}{lccccc}
\hline
Quantity                      &  Notation       & Unit      &  Algol A        &  Algol B        &  Algol C        \\
\hline
Mass                          & $M$             & \Msun     & $3.39\pm0.06$   & $0.770\pm0.009$ & $1.58\pm0.09$   \\
Radius                        & $R$             & \Rsun     & $2.87\pm0.04$   & $3.43\pm0.01$   & $1.7\pm0.3$     \\
Surface gravity               & $\log g$        & [cgs]     & $4.05\pm0.01$   & $3.254\pm0.006$ & $4.18\pm0.16$   \\
Effective temperature         & $T_{\rm eff}$   & K         & $12\,550\pm120$ & $4900\pm300$    & $7550\pm250$    \\
Projected rotational velocity & $v\sin i$       & \kms      & $50.5\pm0.8$    & $62\pm2$        & $12.4\pm0.6$    \\
Synchronous velocity          & $v_{\rm synch}$ & \kms      & $51.2\pm0.5$    & $64\pm1$        & $14.1\pm2.5$    \\
Microturbulence velocity      & $\xi_t$         & \kms      & $0.4\pm0.2$     &                 & $1.68\pm0.06$   \\
Mean metal abundance          & [M/H]           &           & $-0.03\pm0.08$   &                 & $0.04\pm0.09$   \\
Bulk metallicity              & $Z$             &           & $0.014\pm0.002$ &
& $0.012\pm0.002$ \\
\hline
\end{tabular}
\end{table*}

%%%%%%%%%%%%%%%%%%%%%%%%%%%%%%%%%%%%%%%%%%%%%%%%%%%%%%%%%%%%%%%%%%%%%%%%%%%%%%%%%%%%%%%%%%%%%%%%%%%%%%%%%%%%%%%%%%%%%%%%%%%%%%%%%%%%%%%%%%%%%%%%%%%%%

\section{Discussion}

\subsection{Fundamental properties of the components of Algol}

With the velocity semiamplitudes determined for all the components of the Algol system, 
we are able to derive their dynamical masses. The orbital inclinations are taken from 
Richards et al.\ (1988) for the inner orbit, $i_{\rm A-B} = 81.4\pm0.2$, and from Baron 
et al.\ (2012) for the outer orbit, $i_{\rm AB-C} = 83.66\pm0.03$. In the latter study 
$i_{\rm A-B}$ was also derived, and was in agreement with Richards et al.\ (1988) but with 
lower accuracy. These inclinations are well constrained from several different studies and 
techniques. We also used the orbital periods from Baron et al.\ (2012): $P_{A-B} = 
2.867328\pm0.00005$\,d and $P_{AB-C} = 680.168\pm0.54$\,d. We find the masses $M_{A} = 
3.39\pm0.06$\Msun, $M_{\rm B} = 0.770\pm0.009$\Msun\ and $M_{\rm C} = 1.58\pm0.09$\Msun. 
From the outer orbit we get the sum of the masses for Algol A and Algol B, $M_{\rm AB} = 
(M_{\rm A}+M_{\rm B})_{\rm outer} = 4.38\pm0.27$\Msun, which is in accordance with the sum 
of the individual masses of Algol A and Algol B from the inner orbit, $(M_{\rm A} + 
M_{\rm B})_{\rm inner} = 4.16\pm0.06$\Msun. Table~5 contains the physical properties of 
all three stars in the Algol system derived in the current work.

The component masses we find are about 8\% smaller than the most commonly quoted values 
(Richards et al.\ 1988). This corroborates the findings of Baron et al.\ (2012), although 
these authors found masses about 15\% small than those from Richards et al.\ (1988). 
Besides improving the mass values we were able to improve the measurement precision 
to 1.8\% for Algol A and 1.2\% for Algol B. That for Algol C is less improved, at 5\%.

Published determinations of the radii of the stars suffer from degeneracy due to the 
third light and partial eclipses (see Sect.\ 5.1). Richards et al.\ (1988) found $R_{\rm A} 
= 2.90\pm0.04$\Rsun, $R_{\rm B} = 3.5\pm0.1$\Rsun\ and $R_{\rm C} = 1.7$\Rsun\ (no error 
given). Baron et al.\ (2012) achieved better than 0.5\,mas spatial resolution in the $H$-band 
with the CHARA interferometer, and unambiguously resolved the three stars. They found angular 
diameters of $\phi_{\rm A} = 0.88\pm0.05$\,mas, $\phi_{\rm B} = 1.12\pm0.07$\,mas and $\phi_{\rm C} 
= 0.56\pm0.10$\,mas. With the parallax of Algol determined by Zavala et al.\ (2010), 
$\pi  = 34.7\pm0.6$\,mas, they found linear radii of $R_{\rm A} = 2.73\pm0.20$\Rsun, 
$R_{\rm B} = 3.48\pm0.28$\Rsun\ and $R_{\rm C} = 1.73\pm0.33$\Rsun. The interferometric 
measurements agree with our results to within their large uncertainties.

The dimensions of the Roche-lobe filling component B are constrained by the mass ratio. 
This gives another way of determining its linear radius and synchronous rotational velocity. 
With our mass ratio for the semidetached pair, $q_{\rm A-B} = 0.227\pm0.005$ we have a relative 
radius of Algol B of $r_{\rm B} = 0.251\pm0.002$ (radius at the sides of the secondary star 
perpendicular to the line connecting the centres of the two stars). The semi-major axis of 
the inner orbit is $a_{\rm A-B} = 13.65\pm0.07$\Rsun. The linear radius is $R_{\rm B} = 
r_{\rm B}\, a_{\rm A-B} = 3.43\pm0.01$\Rsun. This is consistent with the Baron et al.\ (2012) 
value. The synchronous rotational velocity for this $R_{\rm B}$ is $v_{\rm synch, B} = 
60.2\pm0.2$\kms. We measured $v_{\rm B} \sin i_{\rm B} = 62\pm2$\kms\ from the spectral 
line broadening, which also supports our estimate of the radius of Algol B.

If we use the condition of synchronous rotation for Algol A, which does not have to be 
fulfilled, then for the measured $v_{\rm A} \sin i_{\rm A} = 50.5\pm0.8$\kms\ we get 
$R_{\rm A} = 2.87\pm0.04$\Rsun. If Algol A has been spun up by mass transfer (cf.\ Packet 1981; 
Dechamps et al.\ 2013) then this would be an upper limit for its radius. An improvement in 
direct interferometric measurements of its angular diameter, and/or revising the light curve 
analysis with new constraints from the spectroscopy presented in this work, might allow the 
radius of this component to be measured to high precision.

\subsection{Chemical composition and evolution of the components}

Predictions of carbon depletion in the atmospheres of mass-transferring systems prompted 
several observational studies. Carbon underabundances relative to solar were found by 
Cugier \& Hardorp (1988) in an analysis of far-UV spectra secured from the IUE satellite. 
The carbon abundance they reported for Algol A, relative to the modern standard solar value, 
is [C/H] $ = -0.32\pm0.20$\,dex. Tomkin et al.\ (1993) studied a sample of Algol-type binaries, 
including Algol itself, and confirmed carbon deficiencies in the whole sample. For Algol A 
they found [C/H] $ = -0.22\pm0.15$ dex using high-resolution observations of the \ion{C}{ii} 
4267\,\AA\ line. The reported value is relative to the average abundance, $\log \epsilon(C) = 
8.28\pm0.21$, they determined for the standard stars. Our result reported in Sect.~5.3, 
[C/H]$_{\rm A} = -0.16\pm0.08$, confirms a small carbon depletion for Algol A, both in terms 
of the abundance determined from the strongest carbon line in its spectrum (\ion{C}{ii} 
4267\,\AA) and the deviation of its EW from a calibration derived from standard late-B 
type stars. Nitrogen lines show a slight overabundance with $\log \epsilon(N) = 7.97\pm0.02$, 
but with the caution that the \ion{N}{ii} lines used are very weak (EW $=$ 1.0--2.7\,m\AA). 
Referring to the standard solar nitrogen abundance (Asplund et al.\ 2009), the 
abundance deviation for Algol A is [N/H]$_{\rm A} = 0.14\pm0.05$.

The ratio between C and N abundances is a sensitive indicator of CNO nucleosynthesis 
and the efficiency of mass-transfer and mixing processes in stellar interiors. For Algol A 
we get (C/N)$_{\rm A} = 2.0\pm0.3$ which, compared to the solar value, (C/N)$_\odot = 4.0\pm0.7$, 
indicates that a change in the C/N ratio in Algol A has been detected. CNO-processed layers 
from Algol B, formerly the more massive of the two inner components, are evidently now exposed 
on the surface of Algol A, the mass-gaining and currently more massive star. In our previous 
study on the hot Algol-type binary system u\,Her (Kolbas et al.\ 2014) we determined C/N $=$ 0.89 
for the mass-gaining component. u\,Her contains components with masses of 7.8 and 2.8\Msun, 
substantially larger than the masses of the inner pair in Algol. As predicted by detailed 
chemical evolution models, Kolbas et al.\ (2014) found a stronger carbon depletion in the 
mass-losing component than in the mass-gaining companion. The progenitor mass of what is now 
the less massive component in u\,Her was estimated from evolutionary model calculations to 
be $M_{\rm donor} \sim 7.2$\Msun, whilst an estimate for the initial mass of Algol B is 
$M_{\rm B, init} = 2.7$\Msun\ (Sarna 1993).

Whilst we succeeded in separating and reconstructing the spectrum of Algol B, detailed analysis 
is premature. The disentangled spectrum still suffers from a low S/N, and the spectral lines 
have high rotational broadening. However, this object dominates the X-ray spectrum of the 
system. Drake (2003) used {\it Chandra} Low Energy Transmission Spectrograph observations 
to determine the abundances of C and N in the corona of Algol B. The analysis was performed 
relative to the `standard' star HR\,1099. These two stars have shown many similarities in 
their X-ray spectra, except for the strengths of C and N lines. For N an enhancement by 
a factor of three compared to the standard star is found, whilst no C lines are detected 
in the Algol B spectrum, indicating a C depletion relative to `standard' by a factor of 
ten or more. Moreover, Drake (2003) found a standard (solar) Fe abundance for Algol B, 
in fine agreement with our findings for Algol A and Algol B.

\section{Conclusion}

 The first systematic \'echelle spectroscopic survey of Algol, a hierarchical triple 
system, has been conducted from 2006 to 2010 and covers the entire optical spectral range. 
The technique of spectral disentangling makes it possible, for the first time, to reveal 
the individual spectra of all three components in Algol. In our \spd\ analysis the orbital 
elements for both orbits were optimised, leading to improved measurements of the masses 
of all three component stars. The uncertainties for the masses of the inner (eclipsing) 
system are now below 2\%.

 A detailed spectroscopic analysis of the disentangled spectra of the individual 
components was undertaken, and yielded an accurate determination of the effective 
temperatures and projected rotational velocities for all three stars, and the photospheric 
elemental abundances and bulk metallicity for Algol A and Algol C. Equally importantly, 
tight constraints on the fractional light cotribution of all three components have been
 derived. The spectroscopic light ratio for the inner partially-eclipsing binary could be
 used to lift the degeneracy in the determination of the radii from light curves of the 
system. This would make the system an excellent test case for theoretical work on modelling 
the evolution of mass-transferring binaries, resulting in an improved comprehension of 
Algol systems.

%%%%%%%%%%%%%%%%%%%%%%%%%%%%%%%%%%%%%%%%%%%%%%%%%%%%%%%%%%%%%%%%%%%%%%%%%%%%%%%%%%%%%%%%%%%%%%%%%%%%%%%%%%%%%%%%%%%%%%%%%%%%%%%%%%%%%%%%%%%%%%%%%%%%%

\section*{Acknowledgments}

We thank the anonymous referee for the constructive comments that have helped
to improve the paper.
KP acknowledges funding from the Croatian Ministery of Science and Education through research 
grant (2007-2013), which also enabled a PhD scholarship to VK, and a Zagreb University Research 
Grant. JS acknowledges financial support from STFC in the form of an Advanced Fellowship. 
AT is Postdoctoral Fellow of the Fund for Scientific Research (FWO), Flanders, Belgium. 
Based on observations made with the Nordic Optical Telescope, operated by the Nordic Optical 
Telescope Scientific Association at the Observatorio del Roque de los Muchachos, La Palma, Spain, 
of the Instituto de Astrofisica de Canarias. Bohyunsan Optical Astronomy Observatory (BOAO) is 
an observing facility of the Korea Astronomy and Space Science Institute (KASI), Daejeon, 
Republic of Korea.

\appendix

 \section{Blaze function correction}

Extracted and wavelength-calibrated \'echelle spectra (Sect.~2) are split into orders, 
each of which has a response function determined by the blaze shape. In each order, the intensity 
of signal smoothly rises from the edges to the centre of frame. It is due to the optical setup 
and characteristics of \'{e}chelle spectrograph, and has sine form (see Barker 1984). 
Blaze function removal from 
extracted spectral orders is a critical step in the reduction of \'echelle spectra. Any systematic 
errors in the blaze function correction produces periodic ripples in the shape of the continuum, 
which in turn affects the depth and thus equivalent widths of spectral lines. A major obstacle to 
removal of the blaze function is the presence of broad spectral lines, in particular Balmer lines 
in late-B and early-A stars. These can be wider than a single spectral order, so determining the 
blaze function using points on the spectral continuum is impossible.

The most widespread method to correct for the blaze function is to divide the spectrum with 
a blaze function obtained from flat-fields (c.f.~Erspamer \& North 2002; \v{S}koda et al.\ 2008,
and references therein). 
However, this often gives a wavy shape in the merged 
spectral orders, which then still have to be locally normalised. For unknown reasons blaze 
functions of stellar and flat-field spectra are not identical, and undesired ripples appear 
and degrade the quality of merged spectra.

One of the most important prerequisites for spectral disentangling is the quality of the 
normalised merged \'{e}chelle spectra. To avoid systematic errors, we remove blaze functions 
in more thorough and interactive way. First, the spectral orders suitable for fitting of the
 blaze function are selected (in Fig.~A1 these are orders \#39, and 42). These are orders without 
broad spectral lines and with a substantial
 amount of continuum. High-degree polynomial functions are fitted by least-squares through selected 
continuum points. Wavelengths of maximum intensity of the blaze functions are also calculated, 
and used for the interpolation of the positions of blaze function head in the remaining orders.
 Blaze functions are now interpolated to these orders, as illustrated the upper panel of Fig.~A1.
 The results of the correction using interpolated blaze functions are shown in the middle panel 
of Fig.~A1. The bottom panel of Fig.~A1 illustrates the quality of interpolation and merging 
between two sequential orders containing a Balmer line. It should be emphasized that overlapping 
parts fall on the edges of spectral orders, and suffer a lower S/N than central parts. Despite 
this, merging is good to much better than 1\%.

%%%\newpage
%
\begin{figure}
\includegraphics[width=8.5cm]{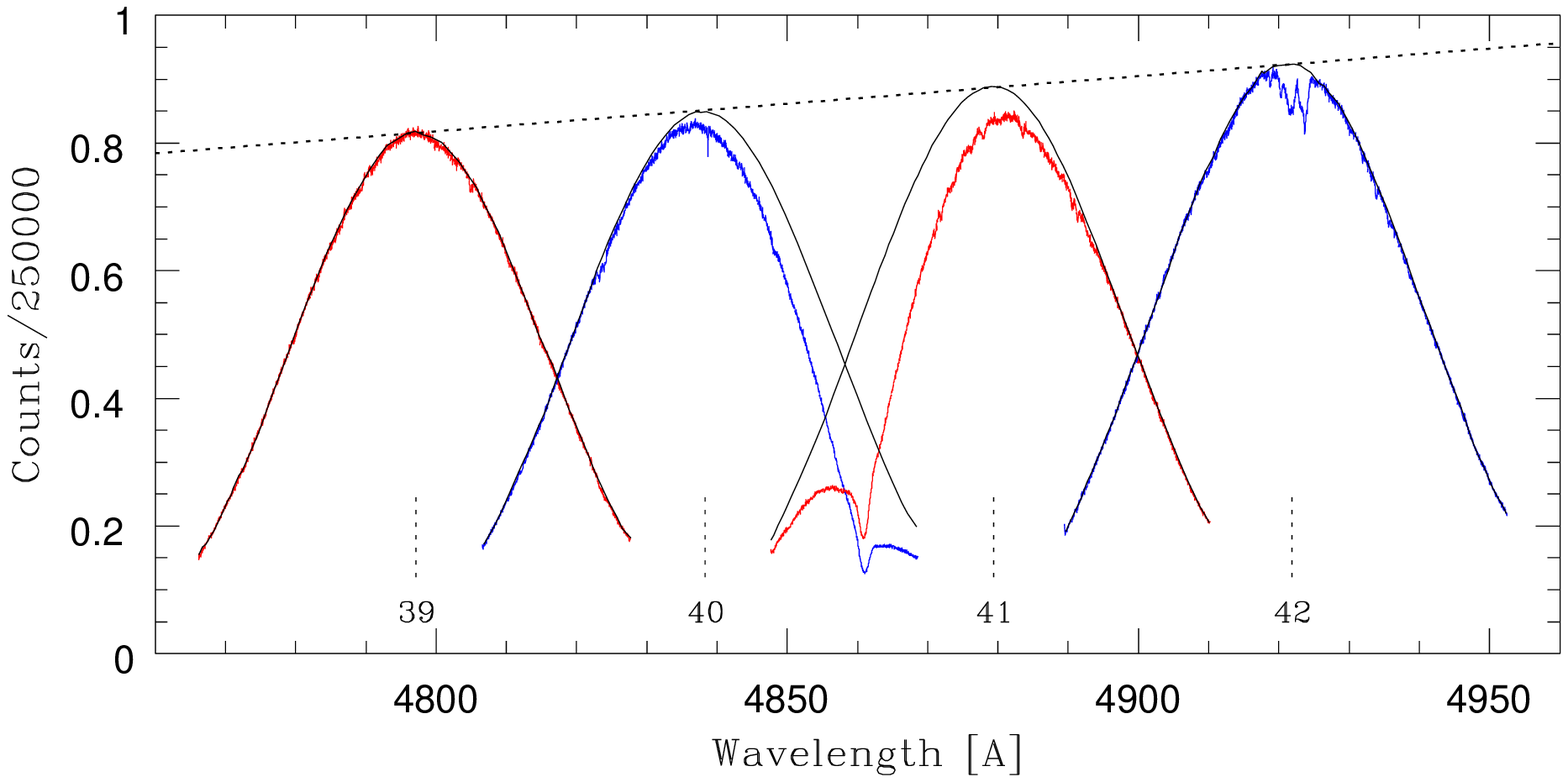}
\includegraphics[width=8.5cm]{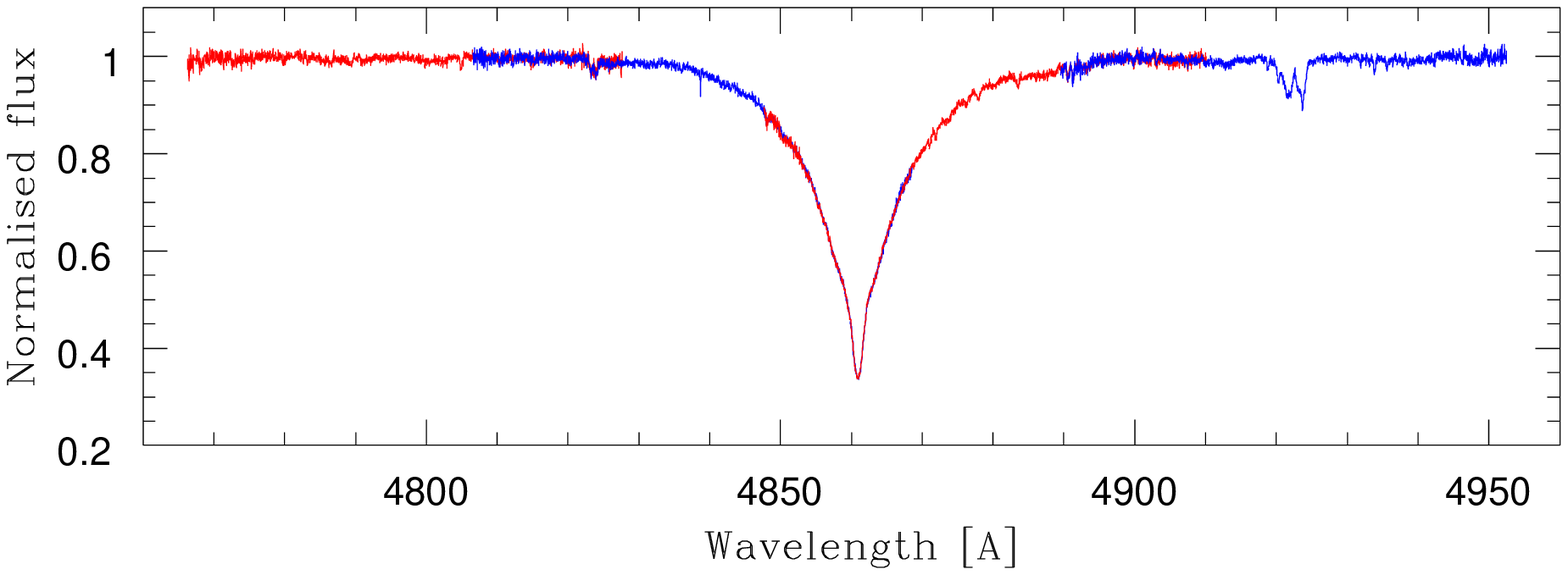}
\includegraphics[width=8.5cm]{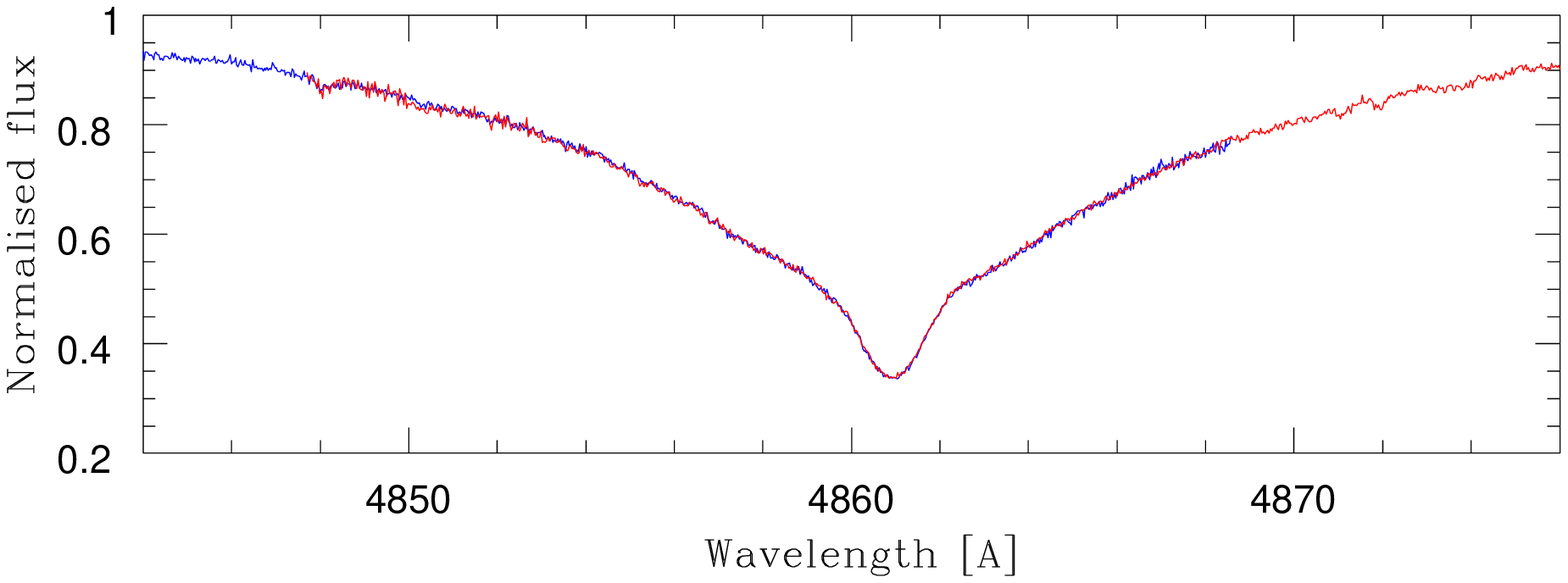}
\vspace{11pt}
\caption{Blaze function correction and merging the spectral orders
contaning broad Balmer lines. From the top to the bottom: Identification
of suitable orders for the determination of the blaze function, and
interpolation to the positions of echell\'{e} orders containing a Balmer
line (upper panel), merging of spectral orders after removing fitted
(orders \#39 and \#42) and interpolated (orders \#40, and \#41) blaze
functions (middle panel), and close up of merged orders in the wings
and core of a Balmer line (bottom panel).}
\label{appenfig}
\end{figure}

%%%%%%%%%%%%%%%%%%%%%%%%%%%%%%%%%%%%%%%%%%%%%%%%%%%%%%%%%%%%%%%%%%%%%%%%%%%%%%%%%%%%%%%%%%%%%%%%%%%%%%%%%%%%%%%%%%%%%%%%%%%%%%%%%%%%%%%%%%%%%%%%%%%%%
\end{document}